\documentclass[aps,pra,twocolumn,superscriptaddress,longbibliography]{revtex4-2}

\usepackage{mathrsfs}
\usepackage{tabularx}
\usepackage{slashed}
\usepackage{amsmath}
\usepackage{amsfonts}
\usepackage{graphicx,xcolor}
\usepackage{times}
\usepackage[caption=false]{subfig}
\usepackage[colorlinks,linkcolor=red,citecolor=blue]{hyperref}
\usepackage{amsthm}

\newtheorem*{presult*}{First Result}

\newtheorem*{mresult*}{Main Result}

\newcommand{\vk}{\mathbf k}

\usepackage{amssymb, amsbsy, latexsym, dsfont, array, layout}
\usepackage{braket}

\begin{document}

\title{Spontaneous Chern-Euler Duality Transitions}

\author{Kang Yang}
\affiliation{\mbox{Dahlem Center for Complex Quantum Systems and Fachbereich Physik, Freie Universit\"at Berlin, 14195 Berlin, Germany}}
\author{Zhi Li}
\affiliation{Perimeter Institute for Theoretical Physics, Waterloo, Ontario N2L 2Y5, Canada}
\author{Peng Xue}
\affiliation{Beijing Computational Science Research Center, Beijing 100193, China}
\author{Emil J. Bergholtz}
\affiliation{Department of Physics, Stockholm University, AlbaNova University Center, 106 91 Stockholm, Sweden}
\author{Piet W. Brouwer}
\affiliation{\mbox{Dahlem Center for Complex Quantum Systems and Fachbereich Physik, Freie Universit\"at Berlin, 14195 Berlin, Germany}}

\date{\today}
\begin{abstract}
Topological phase transitions are typically characterized by abrupt changes in a quantized invariant. Here we report a contrasting paradigm in spontaneous parity-time symmetry-breaking transitions induced by non-Hermiticity. The topological invariant remains conserved, but its nature transitions between the Chern number, characteristic of chiral transport in complex bands, and the Euler number, which characterizes the number of nodal points in pairs of real bands. This conserved morphing of topological properties applies to all pairs of bands whose spectral gaps to other bands remain open during the transition.  It features qualitative changes in the non-Abelian
geometric phases, which quantize in different forms across the transition. Our findings establish a generic topological duality principle governing transitions across symmetry classes and reveal unique connections among topology, symmetry, and non-Abelian gauge structure in non-unitary processes.
\end{abstract}

\maketitle
Symmetry, geometry, and topology play fundamental roles in modern physics, from elementary particles and gravity to quantum materials. Knowledge of the global topology of energy bands in quantum materials provides insights into the nature of the quantum states and gives rise to precisely quantized observables \cite{RevModPhys.89.040502}. The best known example of such a topological invariant is the Chern number, which describes the topology of complex vector bundles in two dimensions. The topological robustness of the Chern number underpins one of the most precisely measured physical constants, the Hall conductance in the quantum Hall effect. A transition between different quantized values requires the closing of a band gap, accompanied by the appearance of Dirac or Weyl points~\cite{RevModPhys.90.015001}.

The advent of dissipative photonic or acoustic systems with parity-time (PT) symmetry has expanded the scope of topological physics~\cite{ep-optics,feng2017non,el2018non,ozdemir2019parity,RevModPhys.93.015005}. PT-symmetric band structures are described by a real-valued ``Hamiltonian'' matrix $H(\vk)$, which is not Hermitian in the presence of loss and gain \cite{RevModPhys.96.045002}. Eigenstates and eigenvalues of non-Hermitian real $H(\vk)$ are either real or appear in complex-conjugated pairs~\cite{Bender_2007}. While $H(\vk)$ itself is PT-symmetric, each individual complex eigenstate that dominates the time evolution breaks PT symmetry, leading to a spontaneous symmetry breaking in the system. Transitions between PT-symmetric and PT-breaking eigenstates take place at exceptional points (EPs) in the spectrum of $H(\vk)$, at which two real eigenvalues of $H(\vk)$ meet and convert into a pair of complex-conjugate eigenvalues~\cite{PhysRevLett.103.093902,PhysRevA.80.052107,ruter2010observation,PhysRevLett.106.093902,hodaei2017enhanced,PhysRevLett.123.213901,tang2020exceptional,PhysRevLett.127.186602,PhysRevLett.127.186601}.

From a topological perspective, real and complex bands of eigenstates are sharply different: Complex bands exhibit chiral transport described by a Chern number, whereas the bands associated with real eigenstates always have a vanishing Chern number and chiral transport is ruled out. On the other side, a pair of real bands may have nonchiral topology described by the integer-valued Euler number~\cite{PhysRevLett.118.056401,PhysRevX.9.021013,bouhon2020non,zhao2020equivariant,PhysRevLett.125.053601,PhysRevLett.125.126403,PhysRevB.103.205303,zhao2022quantum}. A nonzero Euler number signals an obstruction for opening a gap between the two bands \cite{PhysRevX.9.021013,bouhon2020non,PhysRevLett.125.053601}. The possibility to realize two different topological paradigms in the same physical system \cite{yang2023homotopy} sets non-Hermitian PT-symmetric systems apart from their counterparts without symmetries \cite{PhysRevLett.120.146402,PhysRevB.101.205417,PhysRevB.103.155129,wang2021topological,patil2022measuring,ding2022non} as well as Hermitian systems. In this work, we show that the spontaneous real-to-complex transitions in PT-symmetric systems obey a universal Chern-Euler duality principle. 
This duality principle results in a {\it spontaneous Chern-Euler duality transition} that applies to all complete spontaneous PT-symmetry breaking between a pair of bands.

\begin{figure}
    \centering
    \includegraphics[width=1.0\linewidth]{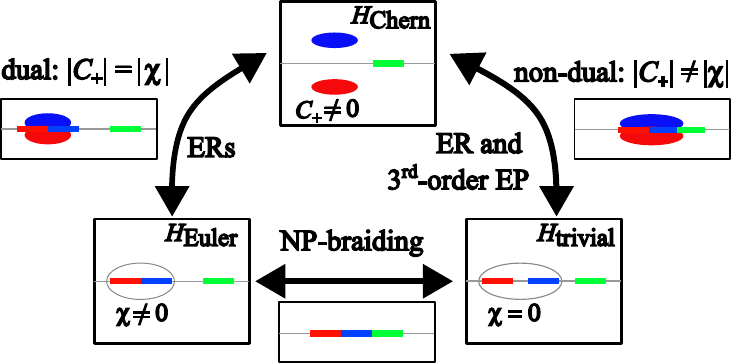}
    \caption{Schematics of the transitions studied in this article. The boxes schematically indicate the spectra at the three types of models $H_{\rm Euler}$, $H_{\rm Chern}$, and $H_{\rm trivial}$ and during the transitions between them. The Chern-Euler duality transition between bands 1 and 2 (shown in red and blue) takes place without involvement of the third real band (green). The non-dual transition between bands with different Chern number and Euler number forces connection among all three bands, a gap closing similar to conventional topological transitions. In the Hermitian regime, the Euler topology requires a minimal number of $2|\chi|$ NPs between the bands 1 and 2 \cite{PhysRevX.9.021013,bouhon2020non,PhysRevLett.125.053601}.
    }
    \label{fig_pd}
\end{figure}

We illustrate this general duality principle by providing three PT-symmetric $3 \times 3$ band structures as minimal examples:
(i) An Hermitian model $H_{\rm Euler}(\vk)$, for which the lowest two bands are separated from the third band by a gap and have nontrivial topology with Euler number $|\chi| = 2$, (ii) a non-Hermitian band structure $H_{\rm Chern}(\vk)$, which has two complex-conjugate complex bands with Chern number $C_{\pm} = \pm 2$ and one real band, and (iii) a Hermitian trivial model $H_{\rm trivial}(\vk)$, which has three trivial bands separated by band gaps. Complex spectra corresponding to these models and their distinct transition mechanisms via nodal points (NPs), exceptional rings (ERs) \cite{zhen2015spawning,Cerjan_2019} and third-order EPs are shown schematically in Fig.\ \ref{fig_pd}.

The real-to-complex transition between $H_{\rm Euler}(\vk)$ and $H_{\rm Chern}(\vk)$ takes place via spontaneous PT symmetry breaking. It starts with the formation of ERs in the lowest two bands. In the course of the transition, the ER sweeps over the Brillouin zone (BZ), so that eventually all eigenvalues and eigenstates in the initial two Euler bands become complex and an imaginary spectral gap opens up. 
The two real bands evolve into two complex bands without connecting to any other bands; the gap to the third (real) band remains open at all times. 
We prove that in general such a ``dual transition'' of a pair of bands can take place {\it if and only if} the absolute value of the Chern number of the complex band pair equals the absolute value of the Euler number of the real band pair,
\begin{equation}
   |C_{\pm}|=|\chi|. \label{eq:EulerChern}
\end{equation}
The ``if'' part may also be realized in explicit symmetry breaking in Hermitian regimes with fine-tuned perturbations \cite{PhysRevLett.122.106405,PhysRevResearch.1.033126}, by extracting a chiral basis in the Euler bands \cite{bouhon2020non,zhao2020equivariant,bouhon2022multi}. The ``only if'' part, the necessary equality, is unique to spontaneous PT-symmetry breaking driven by loss and gain. It mandates two consequences: (I) Chern bands must appear after a spontaneous PT-symmetry breaking inside Euler bands. (II)
If Eq.~\eqref{eq:EulerChern} is not met, including the case where no Euler number can be defined for the real band pair, a spontaneous PT-breaking transition involving only two bands is impossible. Instead, at least three bands must connect during the transition.  Such a ``non-dual'' transition may involve a third-order EP, as we observe for the transition $H_{\rm trivial}(\vk) \to H_{\rm Chern}(\vk)$, see Fig.\ \ref{fig_pd}. 

\emph{Models---}The PT-symmetric three-band models are described by a three-component vector $|\psi(t,\vk) \rangle$ that evolves according to the Schr\"odinger-like equation $i\partial_t|\psi(t,\mathbf k)\rangle=H(\mathbf k)|\psi(t,\mathbf k)\rangle$, where $H(\mathbf k)$ is a real $3\times 3$ matrix periodic in $\mathbf k$, as required by PT symmetry. The two variables $\mathbf k=(k_x,k_y)$, which may be quasi-momenta of photons or tuning parameters of the system \cite{ep-optics,feng2017non,el2018non,ozdemir2019parity}, reside in a torus BZ. The eigenvalues $\omega_j(\vk)$ $(j=1,2,3)$, of $H(\vk)$ form three energy bands, which may be complex if $H(\vk)$ is non-Hermitian. 
A minimum of three bands is required for a non-vanishing Euler/Chern topology in PT-symmetric systems \cite{yang2023homotopy}.

We first summarize the relevant properties of the three reference band structures $H_{\rm Euler}(\vk)$, $H_{\rm Chern}(\vk)$, and $H_{\rm trivial}(\vk)$. Explicit expressions are given in the Appendix A.
For the Hermitian models $H_{\rm Euler}(\vk)$ and $H_{\rm trivial}(\vk)$, all three eigenvalues and eigenstates are real. Both models have a spectral gap between the middle and upper band. We label the eigenvalues as $\omega_1(\vk) \le \omega_2(\vk) < \omega_3(\vk)$.
For $H_{\rm trivial}(\vk)$ there is also a spectral gap between bands 1 and 2, and all the three bands are topologically trivial. 
For $H_{\rm Euler}(\vk)$, the topology of the combined bands $1$ and $2$ is described by the Euler number $|\chi| = 2$. Note that the Euler number is even for a three-band periodic $H(\mathbf k)$ on a torus BZ~\cite{PhysRevB.102.115135,bouhon2022multi,PhysRevB.108.155137,d85h-jq67}. 

The Euler number $\chi$ can be calculated from the non-Abelian Berry curvature $B(\vk)$, a $2 \times 2$ matrix generalizing the Berry curvature to a pair of bands~\cite{PhysRevLett.52.2111},
\begin{equation}
  B(\mathbf k)=\partial_{k_x} A_{y}(\mathbf k)-\partial_{k_y} A_{x}(\mathbf k)-i[A_x(\mathbf k), A_y(\mathbf k)].\label{eq_mbc}
\end{equation}
Here $A_{x/y}(\mathbf k)_{ij}= i\langle \psi_i(\mathbf k)|\partial_{k_{x/y}}\psi_j(\mathbf k)\rangle$ is the non-Abelian Berry connection, matrix indices $i,j\in \{1,2\}$ correspond to bands indices,
and $[\cdot,\cdot]$ denotes the commutator. If $H(\vk)$ is Hermitian, $B(\vk)$ is an antisymmetric $2 \times 2$ matrix and $\chi$ is given by the integral of the off-diagonal component $B_{12}(\mathbf k)$ over the BZ \cite{milnor1974characteristic},
\begin{equation}
    \chi=\frac{1}{2\pi i} \int B_{12}(\mathbf k)d^2k. \label{eq_eu}
\end{equation} 
The expression also works for non-Hermitian real bands after applying an orthonormalization as described in Appendix B.
The Euler number has a sign ambiguity from the gauge choice of the eigenstate basis $|\psi_{1,2}(\vk)\rangle$. Only its absolute value $|\chi|$ appears in physical expressions. 

The continuous transition mechanism of the two Hermitian models $H_{\rm trivial}(\vk)$ and $H_{\rm Euler}(\vk)$ has previously been studied in the literature \cite{PhysRevX.9.021013,bouhon2020non,PhysRevLett.125.053601,wu2019non}. The nonzero Euler number obstructs a direct gap opening between bands 1 and 2 of $H_{\rm Euler}(\vk)$. Instead, the gap opening between these two bands requires a closing of the gap with the third band first. After this gap closing, NPs connecting bands 2 and 3 appear. Winding of these ``2-3 nodes'' relative to the ``1-2 nodes'' between bands 1 and 2 changes the topology of the latter \cite{PhysRevX.9.021013,bouhon2020non,wu2019non}, allowing for the opening of a gap between bands 1 and 2 after the Euler topology is trivialized (see Fig.~\ref{fig_pd}).

In the remainder of this article, we illustrate transitions obtained by adding the anti-Hermitian model $H_{\rm Chern}$ to the Hermitian models, which induces complex eigenstates that spontaneously break the PT symmetry. The Hamiltonian $H_{\rm Chern}$ has two complex bands that are complex conjugates of each other, and one real band. The complex bands are separated by an {\em imaginary} spectral gap, indicating a complete spontaneous PT-symmetry breaking \cite{Bender_2007}. We use $|\phi_+(\mathbf k)\rangle$ to represent the complex eigenstate with eigenvalue $\omega(\mathbf k)$ on the lower half complex plane, and $|\phi_-(\mathbf k)\rangle\equiv|\phi_+(\mathbf k)\rangle^\ast$ to represent the conjugate partner with eigenvalue $\omega^\ast(\mathbf k)$. 
The topology of the complex bands is characterized by the Chern number $C_+ = -C_-$ \cite{yang2023homotopy}, which is the integral of the diagonal Berry curvature $B_{jj}(\vk)$ associated with one of the complex bands 
over the BZ ($j \in \{+,-\}$),
\begin{equation}
  C_j= \frac{1}{2 \pi} \int B_{jj} (\mathbf k)d^2 k. \label{eq:Chern}
\end{equation}

\begin{figure*}
    \centering
    \includegraphics[width=\linewidth]{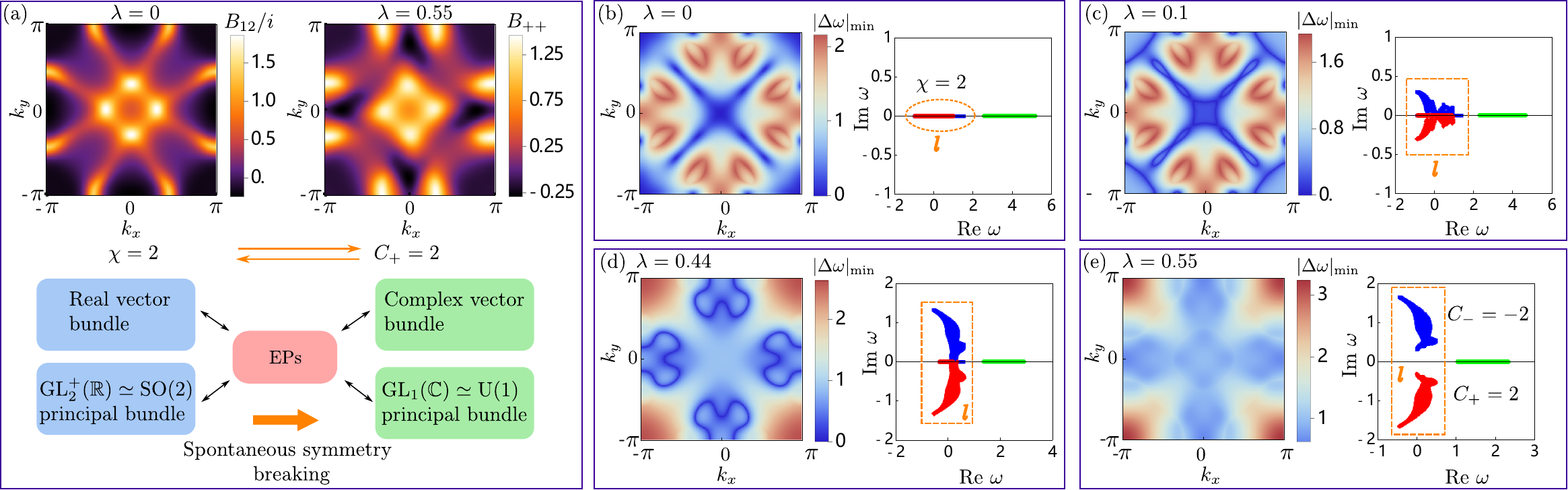}
     \caption{ The spontaneous Chern-Euler duality transition.
    (a) Top: The off-diagonal Berry curvature $B_{12}$ at $\lambda=0$ (left) and  the diagonal Berry curvature $B_{++}$ at $\lambda=0.55$ (right) over the two-dimensional BZ. Their integrals are equal to $\chi=2$ and $C_+ =2$, respectively. Bottom: In all PT-breaking transitions without remote band participation, through EPs, the real bundle spanned by the two real bands with $\mathrm{GL}^+_2(\mathbb{R})$ basis transformation continuously rotates to the realification of the complex conjugate bundles corresponding to the complex bands with $\mathrm{GL}_1(\mathbb{C})$ basis transformation, obeying a general conservation $|\chi|=|C_\pm|$. (b)-(e) The minimum of the local eigenvalue gaps $|\Delta\omega(\mathbf k)|_{\text{min}}\equiv\textrm{Min}_{i,j}|\omega_i(\mathbf k)-\omega_j(\mathbf k)|$ of $H_{\lambda}$ at $\lambda=0.1$ (c), $\lambda=0.44$ (d), and $\lambda=0.55$ (e), reflecting the shapes of level crossings in the BZ. The corresponding complex spectra are shown in the right panel of each figure, with the three bands labeled by different colors.}
    \label{fig_ect}
\end{figure*}

\emph{Chern-Euler duality transition---}We first look at a transition that satisfies Eq.~\eqref{eq:EulerChern}. This is exemplified by interpolations $H_{\lambda}(\vk)$ between $H_{\rm Euler}(\vk)$ and $H_{\rm Chern}(\vk)$, $0 \le \lambda \le 1$,
\begin{equation}
  H_{\lambda}(\vk) = (1-\lambda) H_{\rm Euler}(\vk) + \lambda H_{\rm Chern}(\vk).
  \label{eq:Hlambda}
\end{equation}
Figure \ref{fig_ect}a compares the off-diagonal Berry curvature $B_{12}(\vk)$ in the Hermitian limit $\lambda = 0$ with the diagonal Berry curvature $B_{++}(\vk)$ for $\lambda = 0.55$, where the PT symmetry is spontaneously broken in the entire BZ. Initially at $\lambda = 0$, the integral of $B_{12}(\vk)$ takes the quantized value $\chi=2$. In the fully PT-broken case at $\lambda = 0.55$, the integral of the diagonal Berry curvature $B_{++}(\vk)$ of the complex band is $C_+=2$ ($C_-=-2$ due to the PT symmetry).
The transition from $H_{\rm Euler}(\vk)$ to $H_{\rm Chern}(\vk)$ takes place through the formation of ERs. As $\lambda$ increases, these ERs grow out of the NPs of $H_{\rm Euler}(\vk)$, then sweep over the BZ, see Figs.\ \ref{fig_ect}b--d, until all eigenvalues of bands 1 and 2 have gone through the real-to-complex transition and an imaginary spectral gap is opened between the resulting two complex bands (Fig.\ \ref{fig_ect}e). A key observation is that the real band 3 remains detached from bands 1 and 2 at all times during the spontaneous PT-breaking transition, as shown in Fig.\ \ref{fig_ect}b--e. The opening of an imaginary gap between bands 1 and 2 without intermediate gap closings to any other band is a particular feature of the non-Hermitian regime, as the nontrivial Euler-number topology of bands 1 and 2 prevents the opening of a real gap between them in the Hermitian PT-symmetric regime (\cite{bouhon2022multi,PhysRevLett.125.053601}). 

\emph{Proof for generic cases---}In the above  transition without remote band touching, the topology $|\chi| = 2$ of the real Euler bands 1 and 2 is passed on to that of the complex Chern bands carrying $C_{\pm} = \pm 2$. We prove that this conservation $ |C_{\pm}|=|\chi|$ is not accidental to our example, but must hold generically for all transitions without participation of remote bands, regardless of the exact Euler/Chern numbers and the number of remote bands in the system. It is guaranteed by the band gaps to remote bands: {\em Spontaneous PT-breaking transitions without involvement of remote bands can occur only between real and complex bands with $|\chi| = |C_{\pm}|$} [Eq.\ (\ref{eq:EulerChern})].
This is the main result of this article. A direct consequence is that two-band models cannot host Chern bands after spontaneously breaking the PT-symmetry \cite{yang2023homotopy,PhysRevB.84.153101}, since the real band pair in a two-band model always has $\chi=0$.

We consider a generic $N\times N$ transition Hamiltonian $H(\mathbf k,\lambda)$ parametrized by the ``transition time'' $0\le\lambda\le1$. At $\lambda=0$, two real bands are band-gapped to remote bands and at $\lambda=1$ they eventually become two complex conjugate bands without touching remote bands during $\lambda\in[0,1]$. For $(\mathbf k,\lambda)$ before the PT breaking, the real eigenstates $|\psi_1\rangle,|\psi_2\rangle$ span a 2D real vector space $V^{(1)}_{\lambda}(\mathbf k)$. After the PT breaking, the two complex conjugate eigenstates can be ``realified'' by recombinations $(|\phi_+\rangle+|\phi_-\rangle)/2,(|\phi_+\rangle-|\phi_-\rangle)/2i$. These two realified vectors also span a 2D vector space $V^{(3)}_{\lambda}(\mathbf k)$ for $\lambda,\mathbf k$ in a range different from those for $V^{(1)}$. The boundaries between the two constructions in the $(\mathbf k,\lambda)$-space are EPs, 
where $|\psi_1\rangle, |\psi_2\rangle$ coalesce into a single eigenstate $|\psi_\ast\rangle$ and then split into a complex conjugate pair. For an EP of eigenvalue $\omega_\ast$, although there is only one eigenvector, we have another generalized eigenvector satisfying $(\omega_\ast-H)^2|\psi'_\ast\rangle=0$ \cite{kato2013perturbation}. The real vectors $|\psi_\ast\rangle, |\psi'_\ast\rangle$ span a 2D real vector space $V^{(2)}_\lambda(\mathbf k)$ in the $(\mathbf k,\lambda)$-range of EPs. The most important step is that putting $V^{(1)},V^{(2)},V^{(3)}$ defined in mutually complementary $(\mathbf k,\lambda)$-range together forms a continuous function of spaces $V_{\lambda}(\mathbf k)$ for all $\lambda\in[0,1],\mathbf k\in \textrm{BZ}$. This observation comes from the connection between the projection operator $P_\textrm{sum}$ to the two bands together and the Hamiltonian $H$ \cite{kato2013perturbation}:
\begin{equation}
    P_\textrm{sum}(\mathbf k,\lambda)=\frac{1}{2\pi i}\oint_l dz[z-H(\mathbf k,\lambda)]^{-1},\label{eq_pjo}
\end{equation}
where $l$ can be any loop that encloses the spectrum of the two bands at $(\mathbf k,\lambda)$ (e.g. Fig.~\ref{fig_ect}b-e). The right hand side is continuous whenever the two bands do not touch remote bands \cite{yang2023homotopy}. So $P_\textrm{sum}(\mathbf k,\lambda)$ is real and continuous for the whole transition. The images of $P_\textrm{sum}$ at different $(\mathbf k,\lambda)$ are exactly the spaces $V^{(1)},V^{(2)},V^{(3)}$ \cite{kato2013perturbation,suppm}. This shows the continuity of $V_{\lambda}(\mathbf k)$. Intuitively, generalized eigenvectors help $V^{(2)}$ glue $V^{(1)}$ and $V^{(3)}$ continuously together.

\begin{figure*}
    \centering
    \includegraphics[width=\linewidth]{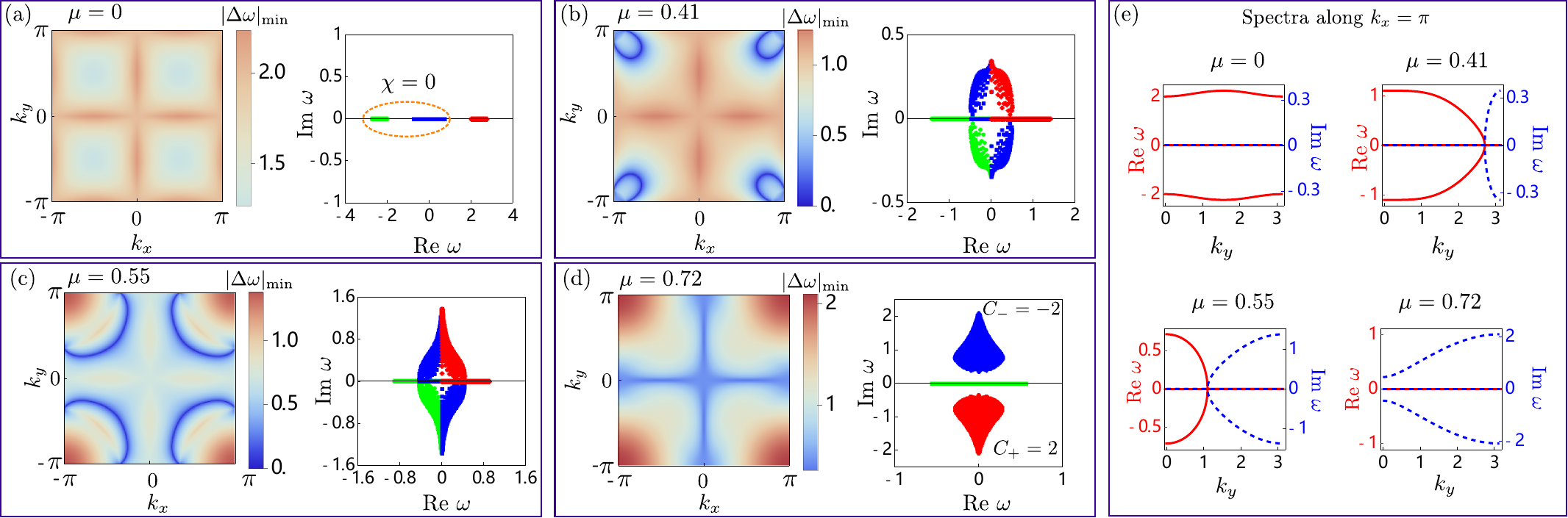}
   \caption{The non-dual transition from trivial bands to Chern bands.
    (a)-(d). The minimum eigenvalue gap $|\Delta\omega(\mathbf k)|_{\text{min}}=\textrm{Min}_{i,j}|\omega_i(\mathbf k)-\omega_j(\mathbf k)|$ and the spectra of $H_{\mu}$ on the complex plane at $\mu=0$ in (a), $\mu=0.41$ in (b), $\mu=0.55$ in (c), and $\mu=0.72$ in (d). All the three bands are connected together during the non-dual transition, which is a universal feature of non-dual transitions.
    (e) The real and imaginary parts of the spectra as a function of $k_y$ at fixed $k_x=\pi$ for different $\mu$. Third-order EPs are generated during this non-dual transition.}
    \label{fig_tct}
\end{figure*}

The continuity of $V_{\lambda}(\mathbf k)$ imposes a rigid connection between the complex bands and the real bands across the transition. Before the transition, the spaces $V_{\lambda=0}(\mathbf k)\equiv V^{(1)}_{\lambda=0}(\mathbf k)$ correspond to the rank-2 real vector bundle $F_{\mathbb R}$ formed by the eigenstates $|\psi_{1,2}(\vk,\lambda=0)\rangle$. After the transition, $|\phi_\pm(\mathbf k)\rangle$ corresponds to complex conjugate rank-1 vector bundles $E_{\mathbb C},E^\ast_{\mathbb C}$ respectively. The two complex bundles admit a  realification $F^{RI}$, given by $V_{\lambda=1}(\mathbf k)\equiv V^{(3)}_{\lambda=1}(\mathbf k)$. The continuity of $V_{\lambda}(\mathbf k)$ tells $F^{RI}\cong F_{\mathbb R}$ \cite{hatcher2003vector}. Note that $F^{RI}$ and $F_{\mathbb R}$ are only isomorphic but usually different.

As a realification, $F^{RI}$ satisfies two properties \cite{milnor1974characteristic,bott1982differential}. First it has trivial first Stiefel-Whitney topology $w_1(F^{RI})=0$, also known as weak topology for real bands \cite{PhysRevLett.116.156402,PhysRevB.95.235425,Ahn_2019,PhysRevB.102.115135}. This means that adiabatic transport of a basis for $F^{RI}$ around the $k_x$ or $k_y$ direction of the BZ-torus preserves its orientation. Second, its Euler number $\chi(F^{RI})$ is related to the Chern number of the complex band $|\chi(F^{RI})|=|C(E_{\mathbb C})|$. Combined with $F^{RI}\cong F_{\mathbb R}$, the first property enforces that if a transition without involving remote bands occurs, an Euler number can be assigned to the initial real band, as this is possible if and only if $w_1(F_\mathbb R)=0$ \cite{milnor1974characteristic}. The second property of $F^{RI}$ dictates $|\chi(F_{\mathbb R})|=|C(E_{\mathbb C})|$. This concludes the generic proof.
A physical illustration is given in Appendix B.

The necessary conservation $ |C_{\pm}|=|\chi|$ is unique to spontaneous PT breaking, which realizes the realification of complex bundles in Chern's construction \cite{Chern}. In comparison, transitions between real and complex bands in Hermitian regimes require explicitly breaking the symmetry in $H(\mathbf k)$ \cite{PhysRevLett.122.106405,bouhon2022multi,PhysRevResearch.1.033126}. In explicit symmetry breaking, the resulting complex bands are not PT-reversal complex conjugate partners. 
The realification cannot be constructed. 
The Chern numbers there are not forced to equal the initial Euler number $\chi(F_{\mathbb R})$. A typical example is the trivial bands $\chi(F_{\mathbb R})=0$. From knowledge of two-band models \cite{RevModPhys.89.040502,PhysRevB.85.165456}, two trivial bands can always be deformed into two Chern bands of $\pm C\ne 0$ without involving remote bands in explicit symmetry breaking. However, this is impossible in spontaneous PT breaking, as shown in the following example of non-dual transitions.

\emph{Non-dual transition---}We contrast the Chern-Euler duality transition with the transition $H_{\rm trivial}(\vk) \to H_{\rm Chern}(\vk)$, which 
is a spontaneous PT-breaking transition involving bands that violate Eq.~\eqref{eq:EulerChern}. We consider the interpolating Hamiltonians
 \begin{equation}
  H_{\mu}(\vk) =(1-\mu) H_{\rm trivial}(\vk) +\mu H_{\rm Chern}(\vk).
\end{equation}
Spectra of $H_{\mu}(\vk)$ at various stages in the transition are shown in Fig.~\ref{fig_tct}a-d. As the real bands in the initial model $H_{\rm trivial}(\vk) $ carry $\chi=0$, the transition to the complex Chern bands of $H_{\rm Chern}(\vk)$ can only take place after all three bands cross simultaneously. The eigenvalues in Fig.\ \ref{fig_tct}b and c indicate the emergence of ERs. The ERs can be put in two categories, those connecting bands $1,2$ and those connecting bands $2,3$. The two types of ERs can intersect at cusps of third-order EPs (Fig.\ \ref{fig_tct}e) from dimension counting \cite{PhysRevLett.127.186602,PhysRevLett.127.186601}, where all three bands become degenerate. The ERs expand and progressively sweep across the BZ as $\mu$ further increases to $0.55$. The transition is complete at $\mu = 0.72$ [Figs.~\ref{fig_tct}d], where two imaginary gaps to bands with Chern number $C_{\pm} = \pm 2$ are opened symmetrically around the real axis. The necessary connection of three bands during all non-dual transitions provides fresh insights on higher-order EPs~\cite{hodaei2017enhanced,PhysRevLett.123.213901,tang2020exceptional}, and serves as a crucial distinction from explicit symmetry breaking transitions.

\emph{Discussion---}In this article, we extended the concept of topological transitions to the realm of topological duality. The spontaneous PT-symmetry breaking in non-Hermitian PT-symmetric systems corresponds to the natural duality between complex vector bundles and real vector bundles. As a result,  the same topology manifests as two distinct quantized properties, depending on the symmetry regime. Dissipative photonic and acoustic systems give this paradigm in vector bundle theory a concrete physical context. 

The Chern-Euler duality transition paves the way for overcoming symmetry constraints in engineering topological states. The dual and non-dual transitions can be diagnosed by measuring different types of ERs in BZ or closed surfaces of parameter spaces on a wide range of platforms \cite{ep-optics,feng2017non,el2018non,ozdemir2019parity,zhen2015spawning,Cerjan_2019,wang2021topological,patil2022measuring,ding2022non}. Berry curvatures are geometric phases around small plaquettes in the $\mathbf k$-space \cite{PhysRevLett.52.2111,DiscreBC}. Due to the imaginary spectral gap, $|\phi_-(\mathbf k)\rangle$ dominates the dynamic evolution. The Berry curvature can be measured by the phase difference between the final states $|\phi_-(\mathbf k)\rangle$ and $|\phi_-(\mathbf k+\delta\mathbf k)\rangle$ at different $\mathbf k$ parameters in few-level-system simulations \cite{wu2019observation,q3y4-3s34}, by performing an evolution $\exp(-iHt)$ with $1/t$ smaller than the imaginary gap for any initial state. Another direct measurement is to match the number of Dirac nodes that reflects the Euler number before the transition and the spectral transport reflecting the Chern number after the transition \cite{yang2026stable}. The latter may be measured by parametric pumping \cite{zilberberg2018photonic}.

The three models are the simplest settings for experimental verification. More illustrating examples are included in the Supplemental Material \cite{suppm}.
In addition to the spectral phenomena presented here, there are many other phenomena related to the topological bands, including various in-gap modes and Wannierization. These properties will be alternative signatures of the transitions that invite future research.

\emph{Acknowledgements---}The authors thank Vatsal Dwivedi and Felix von Oppen for fruitful discussions. KY is supported by the ANR-DFG project (TWISTGRAPH). 
ZL is supported by Perimeter Institute; research at Perimeter Institute is supported in part by the Government of Canada through the Department of Innovation, Science and Economic Development and by the Province of Ontario through the Ministry of Colleges and Universities.
PX is supported by the National Key R\&D
Program of China (Grant No. 2023YFA1406701) and National Natural
Science Foundation of China (Grants No. 12025401, No. 92265209).
EJB is supported by the  Wallenberg Scholars program (2023.0256) and the G\"oran Gustafsson Foundation for Research in Natural Sciences and Medicine.

\bibliography{spnChern}

\subsection*{Appendix A: Reference models}
The reference models $H_{\rm Euler}(\vk)$, $H_{\rm Chern}(\vk)$, and $H_{\rm trivial}(\vk)$ are described by the $3 \times 3$ matrices
\begin{align}
  H_{\rm Euler}(\mathbf k)=&\,
    \begin{pmatrix}
 g_1(k_x,k_y) & f_1(\mathbf k) & f_2(k_x)\\
 f_1(\mathbf k) & g_1(k_y,k_x) & f_2(k_y)\\
 f_2(k_x) & f_2(k_y) & g_2(\mathbf k)
    \end{pmatrix},\label{eq_heu} \\
  H_{\rm Chern}(\vk) =&\,
  \begin{pmatrix}
    0 & h(\vk) & \sin k_y \\
    -h(\vk) & 0 & - \sin k_x \\
    - \sin k_y & \sin k_x & 0
  \end{pmatrix}, \\
  H_{\rm trivial}(\vk) =&\,
  \begin{pmatrix}
    0 & 2 & \sin k_y \\
    2 & 0 & - \sin k_x \\
    \sin k_y & - \sin k_x & 0
  \end{pmatrix},
\end{align}
where $g_1(k_x,k_y)=1+\cos k_y-\cos 2k_x$
, $g_2(\mathbf k)=2-\gamma(\cos k_x+\cos k_y)+\cos 2k_x+\cos 2k_y$, $f_1(\mathbf k)=\cos(k_x-k_y)-\cos(k_x+k_y)$, $f_2(k)=-\gamma\sin k+\sin 2k$, and $h(\vk) = 1 - \cos k_x - \cos k_y$. 
We set $\gamma = 0.6$ to maximize the real spectral gap
between band 3 and bands 1,2 in $H_{\rm Euler}(\vk)$.

\subsection*{Appendix B: Physical illustration of the Chern-Euler Duality Principle}

We assume a generic family of $N$-band PT-symmetric model $H(\vk,\lambda)$ for which a band pair goes through a spontaneous PT-breaking transition parametrized by the ''transition time'' $0 \le \lambda \le 1$. In the Hermitian regime $\lambda=0$, two consecutive real bands in the system are directly band-gapped to other bands, i.e., $\Delta(\mathbf k,\lambda=0)=\textrm{inf }|\omega_i(\mathbf k,\lambda=0)-\omega_j(\mathbf k,\lambda=0)|>0,\forall \mathbf k$, where $i\in\{1,2\},j\not\in \{1,2\}$. The two bands may be topological or not. By including non-Hermitian PT-symmetric couplings, they progressively transition into complex conjugate eigenstates via EPs and  eventually become two complete complex bands separated by an imaginary gap at $\lambda=1$. We show that this process can take place without locally touching other bands, i.e. $\Delta(\mathbf k,\lambda)>0$ in the whole process, if and only if: (a) the real band pair has trivial weak topology (orientable); (b) the Chern numbers of the complex bands equal the Euler number of the real band pair in absolute value. Condition (a) is the prerequisite of an Euler number \cite{milnor1974characteristic}.

{\em  ``If" part---}The ``if'' part is immediate and only slightly different from  Hermitian studies \cite{bouhon2020non,zhao2020equivariant,bouhon2022multi}. When the real bands are orientable, they carry an Euler number \cite{milnor1974characteristic}. For Euler bands $|\psi_{1,2}(\mathbf k)\rangle$ of a Hermitian PT-symmetric Hamiltonian $H(\vk)$, one may construct a pair of chiral basis $|\varphi_\pm(\mathbf k)\rangle=[|\psi_1(\mathbf k)\rangle\pm i|\psi_2(\mathbf k)\rangle]/\sqrt{2}$. Here, the states $|\varphi_\pm(\mathbf k)\rangle$ are \emph{not} eigenstates of $H(\vk)$. Such a transformation diagonalizes the non-Abelian Berry curvature matrix $B(\vk)$,
\begin{equation}
  B'_{++}(\vk) = - B'_{--}(\vk) = i B_{12}(\vk). \label{eq:BB}
\end{equation}
By comparing Eqs.\ (\ref{eq_eu}) and (\ref{eq:Chern}), it immediately follows that the Chern numbers $C'_{\pm}$ of the basis $|\varphi_\pm(\mathbf k)\rangle$ are equal to the Euler number $\chi$ of the real band pair spanned by $|\psi_{1,2}(\mathbf k)\rangle$ in absolute value. 

However, at this point, $|\varphi_\pm(\mathbf k)\rangle$ are only basis transformation; they are not eigenstates of $H(\vk)$ and cannot be distinguished due to their equal energy expectation value [$\textrm{Re }\langle H(\mathbf k)\rangle$] protected by the PT symmetry. 
To realize them as non-degenerate bands of a PT-symmetric $H(\mathbf k)$, we take the following steps. 
First, we flatten the band pair energy. Then we switch on an anti-Hermitian term $i|\varphi_+(\mathbf k)\rangle\langle\varphi_+(\mathbf k)|-i|\varphi_-(\mathbf k)\rangle\langle\varphi_-(\mathbf k)|$. This process does not close the gaps $\Delta(\mathbf k,\lambda)$ to remote bands, hence it is a dual transition.

The above procedure demonstrates the ``if'' part of Eq.~\eqref{eq:EulerChern}. 
However, for generic spontaneous PT-breaking transitions, the initial real bands do not split into the chiral basis. Spontaneous PT-breaking transitions involve ERs sweeping across the BZ, where the eigenvalues coalesce and only a single eigenvector remains \cite{Graefe_2008,hu2023non}. Moreover, away from the ERs, the eigenvectors may deviate significantly from the original pair of real orthonormal eigenvectors $|\psi_{1,2}(\mathbf k)\rangle$ of the Hermitian starting point. Consequently, the resulting complex bands strongly deviate from the chiral basis $|\varphi_{\pm}(\mathbf k)\rangle$. The ``only if'' part of the Chern-Euler duality rule tells how much this difference is. This is answered by realifying complex vectors, an idea initially appeared in Chern's definition of Chern classes~\cite{Chern}.

{\em Real vector bundle structures throughout spontaneous PT breakings.---}In the Hermtian limit, a pair of real eigenstates $|\psi_{1,2}(\vk,\lambda=0)\rangle$ of a PT-symmetric $N \times N$ matrix $H(\vk,\lambda=0)$ spans a two-dimensional subspace $V_{\lambda=0}(\vk) \subset \mathbb R^N$. These subspaces define a real rank-2 vector bundle $F_{\mathbb R,\lambda=0}$.
The key insight in the main text is that this real rank-2 bundle can be extended to the whole transition, irrespective of whether we have two real eigenstates only, band degeneracy at EPs, or a pair of complex-conjugate eigenstates. This leads to a bundle $F_{\mathbb R,\lambda}$ over $(\mathbf k,\lambda)\in\textrm{BZ}\times [0,1]$.

The bundle $F_{\mathbb R,\lambda}$ is constructed as follows. For non-degenerate real eigenvalues (i.e., away from the ERs), one simply builds the vector spaces $V_\lambda(\vk)$ from the two right-eigenvectors $|\psi_{1,2}(\vk,\lambda)\rangle$ of the band pair, the same way as it is done in the Hermitian case. 
For complex conjugate eigenstates $|\phi_{+}(\vk,\lambda)\rangle = |\phi_{-}(\vk,\lambda)\rangle^* \equiv|\phi(\vk,\lambda)\rangle $ one uses $|\psi_1(\vk,\lambda)\rangle=[|\phi_{+}(\vk,\lambda)\rangle+|\phi_{-}(\vk,\lambda)\rangle]/2 =\mbox{Re}\, |\phi_{+}(\vk,\lambda)\rangle$ and $|\psi_2(\vk,\lambda)\rangle = [|\phi_{+}(\vk,\lambda)\rangle-|\phi_{-}(\vk,\lambda)\rangle]/2i=\mbox{Im}\, |\phi_{+}(\vk,\lambda)\rangle$ to span $V_\lambda(\vk)$. This prescription is independent of the phase of $|\phi_{+}(\vk,\lambda)\rangle$, as a different phase choice is a $\mbox{GL}_2(\mathbb R)$ transformation within $V_\lambda(\vk)$. 
At the ER, the eigenvalues of the band pair coalesce to a single real eigenvalue $\omega(\vk,\lambda)$. Although the two bands have only a single right-eigenvector $|\psi(\mathbf k,\lambda)$, one can employ the {\em generalized} right-eigenvector $|\psi'(\mathbf k,\lambda)$, which satisfies the Jordan block relation $|\psi(\mathbf k,\lambda)=[\omega(\vk,\lambda)-H(\vk,\lambda)]|\psi'(\mathbf k,\lambda\rangle$ \cite{kato2013perturbation}, to construct a 2D vector space $V_\lambda(\vk)=\textrm{span}|\{|\psi(\mathbf k,\lambda),|\psi'(\mathbf k,\lambda)\}$ at the ER.
Although $V_\lambda(\vk) \subset \mathbb{R}^N$ is constructed through three distinct procedures for different regimes, it actually varies continuously with $\vk$ across the entire BZ and whole $\lambda\in[0,1]$, due to the band gaps $\Delta(\mathbf k,\lambda)$ to remote bands [cf. Eq.~\eqref{eq_pjo}]. The construction leads to a real rank-2 vector bundle $F_{\mathbb R,\lambda}$ over $(\mathbf k,\lambda)\in\textrm{BZ}\times [0,1]$.

{\em ``Only if'' part---}
Since $F_{{\mathbb R},\lambda=0}$ and $F_{{\mathbb R},\lambda=1}$ are the boundaries of the global bundle $F_{\mathbb R,\lambda}$, they are isomorphic to each other $F_{{\mathbb R},\lambda=1}\cong F_{{\mathbb R},\lambda=0}$ \cite{hatcher2003vector}. Note that isomorphism means that there exists a continuous $N\times N$ rotation $U(\mathbf k)$ in $\mathbb R^N$  (not $2\times2$ rotation) such that $V_1(\mathbf k)=U(\mathbf k)V_0(\mathbf k)$. As these are $N\times N$ rotations, $V_0(\mathbf k)$ and $V_1(\mathbf k)$ are different 2D planes in $\mathbb R^N$. The complex bands obtained via ERs are thus very different from the chiral basis, in accordance with our discussion in the ``if'' part.  An intuitive way to understand the isomorphism is to regard $V_1(\vk)$ as a function continuously evolving from $V_0(\vk)$ via $V_\lambda(\vk)$. Continuous evolution does not alter the global topology in the $\mathbf k$-space. 

We then give a physical illustration of the two properties of realification in the main text, which are knowledge from vector bundle theory \cite{milnor1974characteristic,bott1982differential}. The first is about trivial weak topology, leading to condition (a). At $\lambda=1$, we have complex conjugate eigenstates $|\phi(\vk,\lambda=1)\rangle= | \psi_1(\vk,\lambda=1)\rangle + i | \psi_2(\vk,\lambda=1)\rangle$ and $| \phi(\vk,\lambda=1)\rangle^* = | \psi_1(\vk,\lambda=1)\rangle - i | \psi_2(\vk,\lambda=1)\rangle$ across the whole BZ. 
These real and imaginary parts $| \psi_1(\vk,\lambda=1)\rangle$ and $| \psi_2(\vk,\lambda=1)\rangle$ span a two-dimensional real vector bundle $F^{RI}=F_{{\mathbb R},\lambda=1}$ over the BZ. A unitary gauge transformation on $| \phi(\vk)\rangle$ --- multiplying it by a phase $e^{i\theta}$ --- is equivalent to acting on its real and imaginary components with a $\mbox{SO(2)}$ matrix. This shows that the gauge transformations on $| \psi_1(\vk,\lambda=1)\rangle$ and $| \psi_2(\vk,\lambda=1)\rangle$ can be reduced from $\mathrm{GL}_2(\mathbb R)$ to $\mathrm{SO}(2)$. The vector bundle spanned by $| \psi_1(\vk,\lambda=1)\rangle$ and $| \psi_2(\vk,\lambda=1)\rangle$ is thus orientable \cite{milnor1974characteristic,bott1982differential}, leading to trivial weak topology at $\lambda=1$: $w_1(F_{{\mathbb R},\lambda=1})=0$. As  $F_{{\mathbb R},\lambda=0}\cong F_{{\mathbb R},\lambda=1}$, this implies the trivial weak topology of $F_{{\mathbb R},\lambda=0}$. An Euler number can be assigned to $F_{{\mathbb R},\lambda=0}$ \cite{milnor1974characteristic}.

Now we demonstrate the second property of realification mentioned in the main text.
The idea is to reverse the procedure preceding Eq.\ (\ref{eq:BB}) to show that $\chi(F_{{\mathbb R},\lambda=1})$ is equal to the Chern numbers $C_{+} = -C_{-}$ of the individual complex bands at $\lambda = 1$.
In the non-Hermitian regime, $| \psi_1(\vk,\lambda)\rangle$ and $| \psi_2(\vk,\lambda)\rangle$ are not orthonormal, which prevents direct application of Eq.\ (\ref{eq:BB}). To express the Euler number \cite{milnor1974characteristic}, we orthonormalize the basis vectors $|\psi_1(\vk,\lambda)\rangle$ and $|\psi_2(\vk,\lambda)\rangle$ \cite{yang2023homotopy}.
We then continuously deform the complex eigenstates $| \phi(\vk,\lambda)\rangle \to |\phi'(\vk,\lambda)\rangle \equiv |\psi'_1(\vk,\lambda)\rangle + i | \psi'_2(\vk,\lambda)\rangle$, such that the real and imaginary parts $|\psi'_1(\vk,\lambda)\rangle$ and $|\psi'_2(\vk,\lambda)\rangle$ of $|\phi'(\vk,\lambda)\rangle$ are orthonormal. 
Explicitly, $| \phi'(\vk,\lambda)\rangle$ may be constructed via the Gram-Schmidt orthogonalization procedure,
\begin{align}
  | \phi'\rangle =&\, \frac{ | \psi_1 \rangle}{\sqrt{n_{11}}}
    + i \frac{ n_{11} | \psi_2\rangle - n_{12} | \psi_1\rangle}{\sqrt{n_{11} \det n}},
\end{align}
where $n_{\alpha \beta}(\vk) = \langle  \psi_{\alpha}(\vk)| \psi_{\beta}(\vk)\rangle$,
whereas the continuous deformation  $|\phi(\vk)\rangle \to | \phi'(\vk)\rangle$ may be accomplished by linear interpolation,
$|\phi_{\mu}\rangle =
  (1-\mu) |\phi\rangle + \mu | \phi'\rangle$.
The complex vectors $|\phi_{\mu}(\vk,\lambda)\rangle$ have linearly independent real and imaginary parts, ensuring a well-defined Chern number $C_{\mu}$ for all $0 \le \mu \le 1$. This Chern number is unchanged since the deformation is continuous $C_{\mu=1}=C_{\mu=0}\equiv C_+$.
Along the deformation, the two-dimensional spaces spanned by $|\psi'_1(\vk)\rangle,|\psi'_2(\vk,\lambda)\rangle$  stay the same as $|\psi_1(\vk)\rangle,|\psi'_2(\vk,\lambda)\rangle$, hence the Euler number remains at $\chi(F_{{\mathbb R},\lambda=1})$. 
For $\mu=1$, real and imaginary parts of $|\phi_1(\vk,\lambda)\rangle = | \phi'(\vk,\lambda)\rangle$ are orthonormal, allowing the application of Eq.\ (\ref{eq:BB}), which gives $|C_{\mu=1}| = |\chi(F_{{\mathbb R},\lambda=1})|$. With $\chi(F_{{\mathbb R},\lambda=0})=\chi(F_{{\mathbb R},\lambda=1})$, we obtain condition (b). 

The construction $|\psi_1\rangle=[|\phi_{+}\rangle+|\phi_{-}\rangle]/2$, $|\psi_2\rangle=[|\phi_{+}\rangle-|\phi_{-}\rangle]/2i$ for complex bands parallels the ``realification'' of a complex line bundle in mathematical literature \cite{milnor1974characteristic,bott1982differential}. The realification of a $N$-dimensional complex vector space is to treat it as a $2N$-dimensional real vector space. For example, $\mathbb C^N$ can be regarded as $\mathbb R^{2N}$ by extracting the real and imaginary components (Sec. 22 of \cite{bott1982differential}). At $\lambda=1$,  the complex eigenstate $|\phi(\vk,\lambda=1)\rangle$ spans a complex line bundle $E_\mathbb C$. By extracting the underlying real structure of the complex vector space at every $\mathbf k$, a rank-2 real bundle $E_\mathbb R$ with basis $|\phi\rangle,i|\phi\rangle$ can be defined and is isomorphic to $F_{\mathbb R,\lambda=1}=F^{RI}$, by mapping $(a+ib)|\phi\rangle\to \textrm{Re }(a+ib)|\phi\rangle=a|\psi_1\rangle-b|\psi_2\rangle$. Here, the $\mathbb R$-linear structure of $E_\mathbb R$  is directly constructed from the $\mathbb C$-linear structure on $E_\mathbb C$ \cite{suppm}.

Conversely, the dual spontaneous PT-breaking transition physically decomposes the complexification of an oriented real rank-2 vector bundle up to isomorphism: With an oriented real rank-2 vector bundle $F_{\mathbb R}$ spanned by (not necessarily orthonormal) basis vectors $|\psi_1(\vk)\rangle$ and $|\psi_2(\vk)\rangle$, one may associate a complex rank-2 bundle $F_{\mathbb C}=\mathbb C\otimes_{\mathbb R} F_\mathbb R$, spanned by $|\psi_1(\vk)\rangle$ and $|\psi_2(\vk)\rangle$ with complex coefficients \cite{bott1982differential}. This complex rank-2 bundle can be decomposed into two complex line bundles, $F_{\mathbb C}\cong E_\mathbb C\oplus E^\ast_\mathbb C$. 
From the proof, the complex bands that appear in PT-symmetric models \emph{must} correspond to this realification-complexification duality up to vector bundle isomorphism. The linear independence between the basis vector $|\phi(\vk)\rangle$ and $|\phi(\vk)\rangle^*$ for all $\vk$ is ensured by the presence of an imaginary gap in the complex spectrum. In comparison, explicit symmetry breaking lacks the crucial realification, so the Chern numbers there can be arbitrary. The missing of realification (necessary $|\chi|=|C_\pm|$) also applies to interaction-induced spontaneous symmetry breaking~\cite{PhysRevLett.124.097601}.

A final remark is that complex eigenvalues may braid with each other and do not permit continuous projection operators periodic in $\mathbf k$ \cite{PhysRevLett.120.146402,yang2023homotopy}. This does not affect the questions here. In our case, the initial band pair does not braid with remote bands in eigenvalues, so $P_{\mathrm{sum}}(\mathbf k,\lambda=0)$ is periodic. Since we assume there is no direct band-gap closing with remote bands in the whole transition process, no braids with remote bands will be introduced \cite{yang2023homotopy}. Thus $P_{\mathrm{sum}}(\mathbf k,\lambda=1)$ remains periodic.

\section*{Supplemental Materials}

\subsection{Berry curvature in non-Hermitian regimes}

We clarify cautions of Berry curvature in non-Hermitian regimes. We denote the connection matrix constructed from right eigenstates as Hermitian connection matrices. Another common choice of connection is constructed from biorthogonal eigenstates \cite{zhao2015robust,PhysRevLett.120.146402,PhysRevLett.121.136802}, which we call the covariant connection. The covariant connection is more difficult to compute or measure as it requires the knowledge of left-eigenstates, too. It has the merit of being $\mbox{GL}_n(\mathbb C)$-covariant (see Supplemental Material Sec.~\ref{sc_cpcn} and Ref.~\onlinecite{yang2023homotopy}), while the Hermitian connection is only $\mbox{U}(n)$-covariant. The two choices of connection matrices give identical topological results \cite{PhysRevLett.120.146402,yang2023homotopy} when the states being studied are orthonormalized. In these situations, the gauge structures are reduced to $\mbox{U}(n)$ groups. The invariant polynomials of the Hermitian and the covariant curvatures are in the same de Rham cohomology class according to the Chern-Weil theory. For non-orthonormal state(s), the  $\mbox{GL}_n(\mathbb C)$-covariance is important and the most relevant cases are when we consider the sum topology of multiple bands that possess EPs or eigenvalue braiding (see Supplemental Material Sec.~\ref{sc_nabc}). In our model, each Chern band is well separated from others by spectral gaps and there is no eigenvalue braiding. After normalizing the right eigenstates, the gauge structure of each Chern band is reduced to $\mbox{U}(1)$. Both the Hermitian and the covariant Berry curvatures quantize to the same Chern number, justifying the usage in the main text. An interesting phenomenon is that the covariant connection matrix exhibits seeming singularities near EPs. A geometric explanation is given in Supplemental Material Sec.~\ref{sc_sqm}.

\subsection{Lift of nodal structure protection in spontaneous symmetry breaking regime}\label{sc_pfps}

The irremovable Dirac points obstruct a real spectral gap (energy gap) within Euler bands \cite{bouhon2022multi,PhysRevB.103.205303,zhao2022quantum,PhysRevLett.125.053601}. We recapitulate this picture and give a general proof based on vector-bundle theory. 

Choosing any region $D$ homeomorphic to a disc in the BZ, a local Euler number $2\pi i\chi_D=\int_D B^{12}_{xy}d^2k-\int_{\partial D}\mathbf A^{12}\cdot d\mathbf k$ is defined by subtracting the boundary contribution on $\partial D$. The nodal points inside $D$ can be brought together to be annihilated only if $\chi_D=0$ \cite{bouhon2020non}. This condition can also be understood as a consequence of winding number compensation. Inside a $D$, the two Euler bands are effectively viewed as a two-band system. A PT-symmetric two-band system is protected by a winding number $W^D$ of $(d_x,d_z)$-vectors \cite{wu2019non,bouhon2020non,PhysRevX.9.021013},  where $d_x,d_z$ are the coefficients of the Pauli matrices $\sigma_x,\sigma_z$. Every Dirac point carries a nontrivial winding number $W^D_j$ and $\chi_D$ is equal to the sum of winding numbers $\chi_D=\sum_{j\in D}W^D_j/2=W_{\partial D}/2$ \cite{PhysRevX.9.021013,bouhon2020non}, where the symbol $W_{\partial D}$ represents the winding number along $\partial D$. If these Dirac points can be annihilated inside $D$, $W_{\partial D}=0$, indicating $\chi_D=0$.
Now we have a pair of Euler bands. We partition our BZ into several discs $D_j$, the Euler number of the system is the summation of all local Euler numbers: $\chi=\sum_j \chi_{D_j}$. If we assume all Dirac points within the Euler bands can be annihilated, we obtain a phase $\chi_{D_j}=0$ for all $j$. This requires $\chi=\sum_j \chi_{D_j}=0$, resulting in a contradiction.

A more rigorous proof for the forbidden energy gap is done using Whitney sum \cite{milnor1974characteristic}. If an energy gap is created within the Euler bands without touching other bands, this pair of Euler bands are decomposed into two subbands. This means the Bloch bundle $F$ of the Euler bands can be written as a sum of two rank-1 Bloch bundles $F=F_1\oplus F_2$. First, assuming $F_1$ and $F_2$ are orientable, then their Euler classes are trivial $e(F_1)=e(F_2)=0$ (the Euler class of an orientable line bundle must vanish). Using the Whitney sum, the Euler class of $F$ must also be trivial $e(F)=e(F_1)\smile e(F_2)=0$. This contradicts the nontrivial Euler number of $F$, which is the integral of $e(F)$. Second, when $F_1$ and $F_2$ are not orientable, this is equivalent to saying their first Stiefel-Whitney classes $w_1(F_j)\in H^1(T^2,\mathbb Z_2)=\mathbb Z_2\oplus \mathbb Z_2$ are nontrivial, where $T^2$ is the 2d BZ torus. Let us consider the four-fold covering map  of $T^2$ by itself $p_{(4)}:T^2\to T^2, (k_x,k_y)\to (2k_x,2k_y)$. The pullback bundle $p_{(4)}^\ast F_j$ must be orientable, since $p_{(4)}^\ast w_1(F_j)=2w_1(F_j)=0$. One the other hand, the pullback bundle $p_{(4)}^*F$ carries nontrivial Euler class $e(p_{(4)}^*F)=4e(F)\ne 0$ in $H^2(T^2,\mathbb Z)=\mathbb Z$. Hence, we are back to the orientable case and still have a contradiction.

In the non-Hermitian regime ERs create the following effects. When non-Hermitian terms are turned on, the Dirac nodes expand into ERs that separate BZ into symmetry-preserving regions and spontaneous symmetry breaking (SSB) regions. In symmetry-preserving regions, we may still define a local winding number of $(d_x,d_z)$-vectors \cite{yang2023homotopy}. However, there is no winding-number protection in SSB regions  \cite{yang2023homotopy}. As ERs percolate and span the BZ, all local winding numbers in symmetry-preserving regions are destroyed and we are left with SSB regions (Fig.~\ref{fig_separagap}b).

Alternatively, from the vector bundle point of view, an energy gap between the two bands is equal to decomposing the real Bloch bundle into two lower-rank real vector bundles, which is obstructed by the Euler number. In contrast, the imaginary spectral gap (dissipation gap) opening is not a vector bundle decomposition. Instead, it gives a spontaneous complex structure to the real vector bundle (see proof of the Chern-Euler duality). This is a novel effect brought by non-Hermiticity. Thus the dissipation gap is not restricted by the Euler number. 

\subsection{Review of homotopy analysis}\label{sc_htpa}

Homotopy analysis gives a complete view of the topology inside our system. A PT symmetric system is to have a real $N\times N$ matrix $H(\mathbf k)$ with $\mathbf k$ being the control parameters or momenta. We will use the idea of band partition \cite{yang2023homotopy} to describe the spectral gaps and topology of the system. 

\begin{figure}
    \centering
    \includegraphics[width=0.8\linewidth]{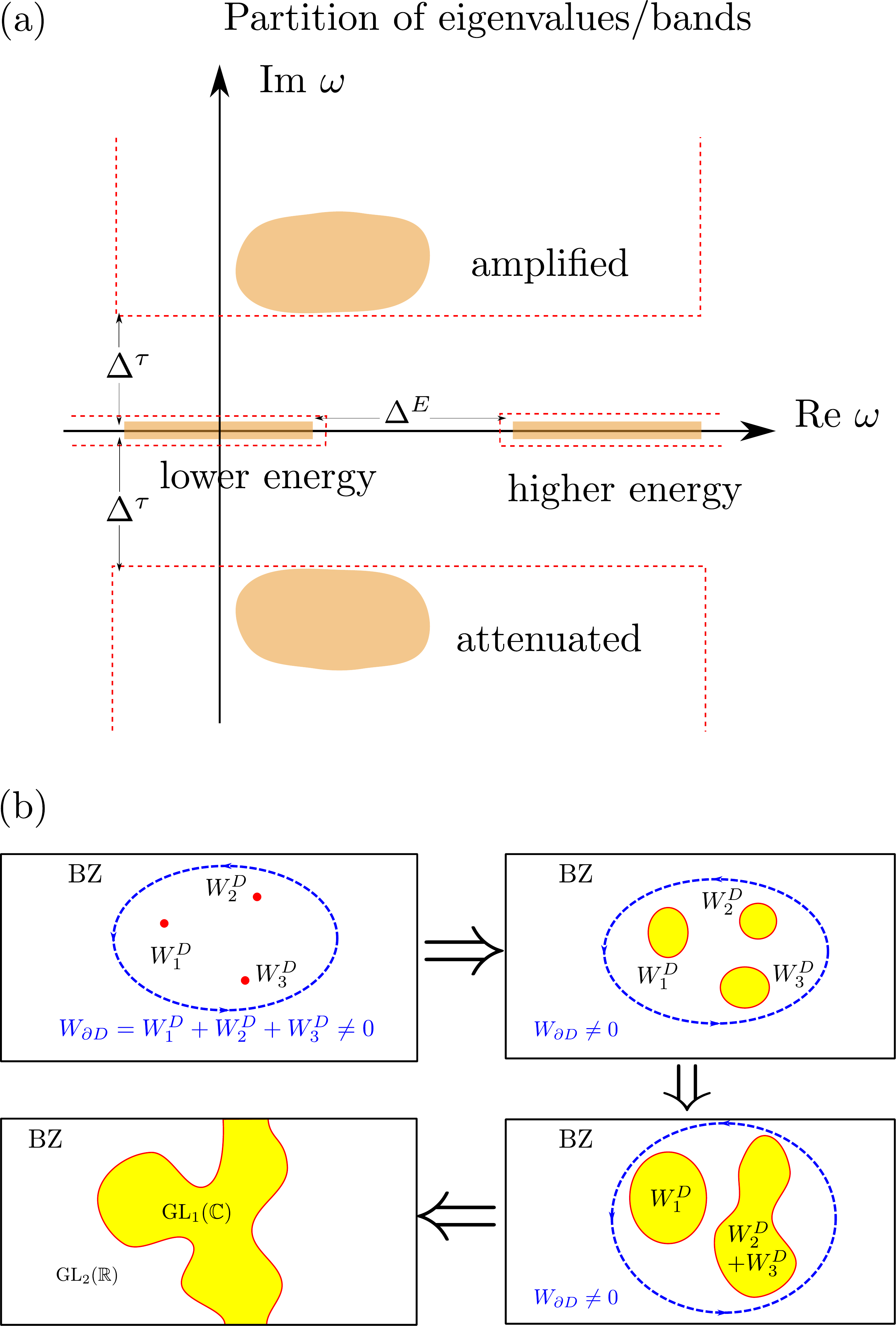}
    \caption{Partition of eigenvalues/bands and spontaneous symmetry breaking (SSB). (a) When the eigenvalues are on the real axis, we may distinguish them by their real parts, and there can be an energy gap between different connected components in the spectra along the real axis. When the eigenvalues can detach from the real axis, we will also have different connected components on the lower and upper half complex planes. They are distinguished by disspation gaps. (b) The SSB inside a BZ. Exceptional rings are the boundaries between symmetry-preserving regions and SSB regions. They convert real gauge structures of the eigenstates into complex gauge structures. The protections of the Dirac points are erased after the gauge type shift.}
    \label{fig_separagap}
\end{figure}

The topology of non-Hermitian PT symmetric systems is captured by the homotopy groups of the matrices having the specified spectral gaps. There are two types of spectral gaps. For those eigenvalues connected to the real axis, we can sort them from lower energy to higher energy. These spectrum can have energy gaps $\Delta^E$ separating lower-energy eigenvalues and higher-energy eigenvalues. Another type of spectral gap comes from comparing the imaginary parts of the eigenvalues. In PT-symmetric systems, the eigenvalues can detouch from the real axis after spontaneous symmetry breaking. The eigenmodes on the lower complex plane are decaying while the eigenmodes on the upper complex plane are amplifying. Together with those oscillating modes on the real axis, the dynamic properties of the system are determined by the spectral gaps $\Delta^\tau$ in the imaginary parts of the eigenmodes (Fig.~\ref{fig_separagap}a). 

From these discussion, we impose the following band/eigenvalue partitions according to the symmetry property of a pair of eigenstates that we study. In the symmetry-preserving regime,  all bands are on the real axis. The band pair we study is assumed to be spectrally gapped to all other bands. We partition the system into two groups: the real band pair and all remote $N-2$  bands. When this band pair goes through a complete spontaneous PT symmetry breaking and detouch from the real axis, we partition the system into three groups: the attenuated complex band of the band pair in the lower half complex plane, the amplified complex band of the band pair in the upper half complex plane, and the remote $N-2$ bands.

When all eigenvalues and eigenvectors are real, the two eigenstate $|\psi_1\rangle,|\psi_2\rangle$ of the band pair and rest of the eigenstate form an $N\times N$ invertible matrix, parametrized by a matrix in $\mbox{GL}_N(\mathbb R)$ (note that if there are exceptional points where some eigenstates coalerce, we need to include the generalized eigenvectors in the Jordan normal form). Meanwhile, there are gauge transformations within the Hilbert space spanned by the band pair and the counterpart spanned by the remote-band eigenstates. A $\mbox{GL}_2(\mathbb R)$ transformation on $|\psi_1\rangle,|\psi_2\rangle$ leaves the Hilbert space $\textrm{span}\{|\psi_1\rangle,|\psi_2\rangle\}$ invariant, while a $\mbox{GL}_{N-2}(\mathbb R)$ transformation on the rest $N-2$ eigenstates leaves the Hilbert space spanned by them invariant . Thus for the eigenstate topology, the Hamiltonian gives rise to a map from the PT preserving regions of BZ, denoted as $\mathrm{BZ}_R$, to the real Grassmanian (up to homotopy equivalence):
\begin{equation}
    f_1:\ \mathrm{BZ}_{R}\to \frac{\mbox{GL}_{N}(\mathbb R)}{\mbox{GL}_2(\mathbb R)\times \mbox{GL}_{N-2}(\mathbb R)}.
\end{equation}
When the eigenvalues of the band pair become a complex conjugate pair, the Hamiltonian gives a map from the PT broken regions of BZ, denoted as $\mathrm{BZ}_C$, to a generalized mixed Grassmanian \cite{yang2023homotopy}:
\begin{equation}
    f_2:\ \mathrm{BZ}_C\to \frac{\mbox{GL}_{N}(\mathbb R)}{\mbox{GL}_1(\mathbb C)\times \mbox{GL}_{N-2}(\mathbb R)}.\label{eq_ssbH}
\end{equation}
This result can be understood as follows: the complex eigenvectors are $|\phi\rangle,|\phi^\ast\rangle$. The real part and the imaginary part of $|\phi\rangle$ are linearly independent vectors in $\mathbb R^N$, to ensure that $|\phi\rangle$ and $|\phi\rangle^\ast$ are linearly independent. Together with a set of real vector basis in the complementary Hilbert space corresponding to the remote bands (this is built similarly as the real space $V(\mathbf k)$ in the Appendix), they from a basis of $\mathbb R^N$, expressed through the column vectors of a real invertible matrix $\mbox{GL}_{N}(\mathbb R)$. However, the system allows for a gauge transformation: $|\phi\rangle\to c|\phi\rangle,\,|\phi\rangle^\ast\to c^\ast|\phi\rangle^\ast, c\in\mathbb C-\{ 0\}$, and a basis transformation in the $(N-2)$-dimensional Hilbert space spanned by the remote bands. These transformations form the group $\mbox{GL}_1(\mathbb C)\times \mbox{GL}_{N-2}(\mathbb R)$. A more systematic proof is to classify the corresponding oblique projection operators into different groups of eigenstates in the band partition, similar to the spectral flattening procedure in~\cite{yang2023homotopy}. Note that due to the coexistence of two types of gauge structures, the spontaneously symmetry breaking regime is usually different from the ten symmetric spaces in Cartan's classification. It is not captured by the K-theory description or the Bott clock when more bands spontaneously break the symmetry.

The lower homotopy groups of the two spaces have been calculated in \cite{yang2023homotopy}. We list out the results:
\begin{align}
    \pi_0\left[\frac{\mbox{GL}_{N}(\mathbb R)}{\mbox{GL}_2(\mathbb R)\times \mbox{GL}_{N-2}(\mathbb R)}\right]=&0,\\
    \pi_1\left[\frac{\mbox{GL}_{N}(\mathbb R)}{\mbox{GL}_2(\mathbb R)\times \mbox{GL}_{N-2}(\mathbb R)}\right]=&\mathbb Z_2,\\
    \pi_2\left[\frac{\mbox{GL}_{N}(\mathbb R)}{\mbox{GL}_2(\mathbb R)\times \mbox{GL}_{N-2}(\mathbb R)}\right]=&\mathbb Z\ (N>4),2\mathbb Z\ (N=3),\nonumber\\ & 2\mathbb Z\oplus 2\mathbb Z\ (N=4),
    \label{eq_hpe}\\
     \pi_0\left[\frac{\mbox{GL}_{N}(\mathbb R)}{\mbox{GL}_1(\mathbb C)\times \mbox{GL}_{N-2}(\mathbb R)}\right]=&0,\\
    \pi_1\left[\frac{\mbox{GL}_{N}(\mathbb R)}{\mbox{GL}_1(\mathbb C)\times \mbox{GL}_{N-2}(\mathbb R)}\right]=&0,\\
    \pi_2\left[\frac{\mbox{GL}_{N}(\mathbb R)}{\mbox{GL}_1(\mathbb C)\times \mbox{GL}_{N-2}(\mathbb R)}\right]=&\mathbb Z\ (N>4),2\mathbb Z\ (N=3),\nonumber\\
    &2\mathbb Z\oplus 2\mathbb Z\ (N=4).\label{eq_hpc}
\end{align}
The Euler number of the band pair in the symmetry-preserving phase comes from Eq.~\eqref{eq_hpe}. For $N=3$ models, the Euler number of the band pair is even. For $N=4$, both the band pair and the remote bands carry Euler numbers that sum to an even integer.
The Chern number in the symmetry-breaking phase comes from Eq.~\eqref{eq_hpc}. For $N=3$ models, the Chern number of each complex band is even. For $N=4$, the remote bands can carry an Euler number. The Chern numbers $C_\pm$ of each complex band in the band pair must have the same parity as the Euler number of the remote bands. The Chern numbers in the spontaneous symmetry breaking regime behave similarly as the Euler number, which agree with the topological duality in the main text. 

Note the quotients in the above expressions are effectively ``gauge structures'' in the eigenstates. This is to say, if we perform these transformations on the eigenstates, the vector space spanned by them is unchanged. In particular, the real and complex general linear groups have the following relation:
\begin{equation}
    \mbox{GL}^+_2(\mathbb R)\simeq \mbox{SO}(2)\cong \mbox{U}(1)\simeq \mbox{GL}_1(\mathbb C),
\end{equation}
where $\simeq$ means homotopy equivalence and $\cong$ means diffeomorphism. The expression $\mbox{GL}^+_2(\mathbb R)$ is the real general linear transformation with a positive determinant. For Euler bands, the gauge structure can indeed be lifted to $\mbox{GL}^+_2(\mathbb R)$ \cite{milnor1974characteristic,bott1982differential} (Euler bands are orientable). In the BZ, ERs are the boundaries separating regions with these two different but homotopy equivalent gauge structures (Fig.~\ref{fig_separagap}b).

\subsection{More mathematical backgrounds of the Chern-Euler duality principle}\label{sc_dpf2}

We first give a simple demonstration for the continuity of the two-dimensional real vector space $V_{\lambda}(\mathbf k)$. Then we give all detailed background knowledge about projection operators, continuity of vector space, and vector bundles for readers not familiar with the mathematical tools. 

\subsubsection{Continuity of \texorpdfstring{$V_\lambda(\mathbf k)$}{V(k)} during the dual transition}

Away from an ER, the right-eigenvectors $|\psi_1(\vk, \lambda)\rangle$ and $|\psi_2(\vk, \lambda)\rangle$ associated with a pair of real eigenvalues $\omega_1(\vk, \lambda)$ and $\omega_2(\vk, \lambda)$ are linear independent. The complex right-eigenvectors $|\phi(\vk, \lambda)\rangle$ and $|\phi(\vk, \lambda)\rangle^*$ of a pair of complex conjugate eigenvalues $\omega(\vk, \lambda)$ and $\omega(\vk, \lambda)^*$ are also linear independent, which means that their real and imaginary parts $|\psi_1(\vk, \lambda)\rangle = \mbox{Re}\, |\phi(\vk, \lambda)\rangle$ and $|\psi_2(\vk, \lambda)\rangle =  \mbox{Im}\, |\phi(\vk, \lambda)\rangle$ are linear independent, too. In both cases, $|\psi_1(\vk, \lambda)\rangle$ and $|\psi_2(\vk, \lambda)\rangle$ span a two-dimensional subspace $V_\lambda(\vk) \subset \mathbb R^N$. These vector spaces form the vector bundle $F_{\mathbb R}$ away from the ERs. For an EP at $\vk = \vk_0,\lambda= \lambda_0$ and eigenvalue $\omega(\vk_0,\lambda_0)$ we use the space spanned by the two {\em generalized right-eigenvectors} to construct the vector space $V_{\lambda_0}(\vk_0)$.

We now show that the assignment of the two-dimensional vector spaces $V_\lambda(\vk) \subset \mathbb R^N$ to a band pair is continuous in the vicinity of an EP. Hereto,
we use that a suitable (non-orthogonal) basis transformation may bring $H(\vk_0,\lambda)$ to the Jordan canonical form
\begin{align}
  &H(\vk_0, \lambda_0) \nonumber\\ = &\begin{pmatrix}
    \omega(\vk_0, \lambda_0)  & 1 & 0 & \ldots & 0 \\
    0 & \omega(\vk_0, \lambda_0) & 0 & \ldots & 0 \\
    0 & 0 & \omega_3(\vk_0,\lambda) & \ldots & 0 \\
    \vdots & \vdots & \vdots & & \vdots \\
    0 & 0 & 0 & \ldots & \omega_n(\vk_0, \lambda)
    \end{pmatrix}.
\end{align}
The space $V_\lambda(\vk_0)$ is spanned by the two generalized eigenvectors at the EP, which are $|1\rangle \equiv (1,0,0,\ldots,0)^{\rm T}$ and $|2\rangle \equiv (0,1,0,\ldots,0)^{\rm T}$. 
% = \mbox{span}(|1\rangle,|2\rangle)$. 
For $\vk$ in the vicinity of the EP, we expand 
\begin{equation}
  H(\vk, \lambda) = H(\vk_0, \lambda_0) + dH,
\end{equation}
where $dH = H'_x dk_x + H'_y dk_y +H'_\lambda d\lambda$ and $d\vk = \vk - \vk_0,d\lambda=\lambda-\lambda_0$. To leading order in $d\vk,d\lambda$, the eigenvalues and eigenvectors of the band pair are solely determined by the single matrix element $dH_{21}$. If $dH_{21} < 0$, there are two real eigenvalues $\omega_{1,2}(\vk, \lambda) = \omega(\vk_0) \pm \sqrt{-dH_{21}}$. If $dH_{21} > 0$, there is a pair of  complex-conjugate eigenvalues $\omega(\vk_0) \pm i \sqrt{dH_{21}}$. In both cases, for small $d\vk,d\lambda$ the left-eigenvectors can be calculated explicitly using degenerate perturbation theory taking into $dH_{21}$ only, followed by non-degenerate perturbation theory to account for the remaining elements of $dH$. Since degenerate perturbation theory amounts to a reorganization of the basis vectors, but otherwise keeps $V_{\lambda_0}(\vk_0) = \mbox{span}(|1\rangle,|2\rangle)$ unchanged, one easily verifies that $V_\lambda(\vk) \to V_{\lambda_0}(\vk_0)$ if $d\vk,d\lambda \to 0$. To leading order in $d\vk,d\lambda$, the Hamiltonian $H(\vk)$ remains a part of the ER if $dH_{21} = 0$. Since the generalized eigenspace is a continuous function of $H(\vk,\lambda)$, in this case, too, one finds that $V_\lambda(\vk) \to V_{\lambda_0}(\vk_0)$ as $d\vk, d\lambda\to 0$. Actually, this continuity is formally defined as a continuous map to the real Grassmannian $\mathrm{Gr}_2(\mathbb R^N)$, as we shall see in the following section.

\subsubsection{More details for projection operators and vector bundles}

We will use the projection operator $P_{\textrm{sum}}(\mathbf k) $ to the band pair to give a formal definition of $V_\lambda(\mathbf k)$ and its continuity in $(\lambda,\mathbf k)$. This will also show that $\chi(F_{\mathbb R,\lambda})$ is a constant in $\lambda$.

The projection into certain eigenspace of $H(\mathbf k)$ has an elegant compact expression \cite{kato2013perturbation,yang2023homotopy}:
\begin{equation}
    \sum_{\omega_j\in \mathcal{D}} P_j=\frac{1}{2\pi i}\int_{\partial\mathcal{D}}(z-H)^{-1}dz,\label{eq_pjo2}
\end{equation}
where $\mathcal D$ is a region homeomorphic to a disc on the complex plane of the spectrum. The resolvent $(z-H)^{-1}$ is the inverse matrix of $(z-H)$, which is well defined when $z$ is not in the spectrum of $H$. 

We need to consider three cases. In the symmetry-preserving regime, where we have real eigenvectors $|\psi_1(\mathbf k,\lambda)\rangle$ and $|\psi_2(\mathbf k,\lambda)\rangle$, $P_{\textrm{sum}}(\mathbf k,\lambda) $ is the sum of the projection operators $P_1(\mathbf k,\lambda)$ and $P_2(\mathbf k,\lambda)$. In the spontaneous symmetry-breaking regime, where we have complex pair eigenvectors $|\phi_+(\mathbf k,\lambda)\rangle=|\phi(\mathbf k,\lambda)\rangle$ and $|\phi_-(\mathbf k,\lambda)\rangle=|\phi^\ast(\mathbf k,\lambda)$, $P_{\textrm{sum}}(\mathbf k,\lambda) $ is the sum of the projection operators $P_+(\mathbf k,\lambda)$ and $P_-(\mathbf k,\lambda)$. When the two eigenvalues are degenerate $\omega_1(\mathbf k,\lambda)= \omega_2(\mathbf k,\lambda)=\omega_\ast$, $P_{\textrm{sum}}(\mathbf k,\lambda) $ is obtained by applying Eq.~\eqref{eq_pjo2} for a small disc containing only $\omega_\ast$.
 
We will show that $P_{\textrm{sum}}(\mathbf k,\lambda) $ is always a real operator, namely, a real $N\times N$ matrix. And its image when acting on $\mathbb R^N$ is exactly $V_\lambda(\mathbf k)$ \cite{kato2013perturbation}, which we will also explicitly present. The continuity of $V_\lambda(\mathbf k)$ results from the continuity of $P_{\textrm{sum}}(\mathbf k,\lambda) $, which is a map to the real Grassmannian $\mathrm{Gr}_2(\mathbb R^N)$. 

We clarify the notations. We consider a system of $N$ bands. Two of its bands are going through a pairwise spontaneous symmetry breaking (SSB). This SSB is achieved through a continuous evolution of Hamiltonian $H(\mathbf k,\lambda)$, where $\lambda$ is the evolution parameter. We assume $0\le\lambda\le 1$ without loss of generality. The initial two bands are band-gapped from other $N-2$ bands, which means that their eigenvalues are not degenerate with any of the other $N-2$ eigenvalues $|\omega_i(\mathbf k)-\omega_j(\mathbf k)|>0,\forall \mathbf k$, where $i\in\{1,2\},j\not\in \{1,2\}$). The transition Hamiltonian is now a map from $\mathrm{BZ}\times [0,1]$ to $N\times N$ real matrices.  For simplicity, we use $\widetilde{\mathrm{BZ}}$ to represent $\mathrm{BZ}\times [0,1]$. The sum projection into these two relevant bands in main result is written as $P_{\textrm{sum}}(\mathbf k,\lambda)$. Before the SSB, we denote the two relevant eigenvalues as $\omega_1(\mathbf k,\lambda)$ and $\omega_2(\mathbf k,\lambda)$ (with corresponding eigenvectors $|\psi_1\rangle$ and $|\psi_2\rangle$). After the SSB, the eigenvalues are denoted as $\omega(\mathbf k,\lambda), \omega^\ast(\mathbf k,\lambda)$ with their corresponding eigenvectors $|\phi_+\rangle=|\phi\rangle, |\phi_-\rangle=|\phi\rangle^\ast$. Their real and imaginary parts are $|\phi_\pm \rangle=|\psi_R\rangle\pm i|\psi_I\rangle$.

We assume that there is no local remote band touching during the transition: $\omega_i(\mathbf k,\lambda)\ne \omega_j(\mathbf k,\lambda)$ for all $i=1,2,\pm$, all $j\ne 1,2,\pm$ and all $\mathbf k,\lambda$. Here we only exclude local band touchings (level crossings at the same $\mathbf k,\lambda$). This is a more general situation where trivial spectral overlaps between bands at different $\mathbf k,\lambda$ are allowed. 

\emph{Images of $P_{\textrm{sum}}(\mathbf k,\lambda)$ in different regimes}: We first verify the relation between the sum projection operator and the real vector bundle structure of rank $2$ in regions of $\widetilde{\mathrm{BZ}}$ without degeneracy between the two relevant bands. We call these regions non-degenerate regions of the $\widetilde{\mathrm{BZ}}$.
In non-degenerate regions, the two eigenvectors are distinct and we can express everything in terms of them.  We denote the regions of $\widetilde{\mathrm{BZ}}$ where the two relevant eigenvalues are real as $\widetilde{\mathrm{BZ}}_R$. The regions of $\widetilde{\mathrm{BZ}}$ where the two relevant eigenvalues are complex as  $\widetilde{\mathrm{BZ}}_C$.  In $\widetilde{\mathrm{BZ}}_R$, we have two real eigenvectors $|\psi_1(\mathbf k,\lambda)\rangle$ and  $|\psi_2(\mathbf k,\lambda)\rangle$. The sum projection into the space $V_\lambda(\mathbb R)$ spanned by $|\psi_1(\mathbf k,\lambda)\rangle$ and  $|\psi_2(\mathbf k,\lambda)\rangle$ is simply given by $P_{\textrm{sum}}(\mathbf k,\lambda)=P_1(\mathbf k,\lambda)+P_2(\mathbf k,\lambda)$, where $P_j(\mathbf k,\lambda)=|\psi_j(\mathbf k,\lambda)\rangle\langle\widetilde\psi_j(\mathbf k,\lambda)|$ and $\langle\widetilde\psi_j|$ are the corresponding biorthogonal left eigenvectors. Note although this is not an orthogonal projection in the presence of non-Hermiticity, $\ker P_{\textrm{sum}}\not\perp \textrm{img } P_{\textrm{sum}}$, the operator $P_{\textrm{sum}}$ still gives us a map from $\widetilde{\mathrm{BZ}}_R$ to the space of two dimensional subspaces of $\mathbb R^N$, the Grassmannian $\mathrm{Gr}_2(\mathbb R^N)$, via:
\begin{equation}
    \widetilde{\mathrm{BZ}}_R\to \mathrm{Gr}_2(\mathbb R^N):\quad  (\mathbf k,\lambda)\mapsto\textrm{img } P_{\textrm{sum}}(\mathbf k,\lambda).
\end{equation}
This is a continuous map on $\widetilde{\mathrm{BZ}}_R$, as the eigenvectors are continuous away from the EP \cite{kato2013perturbation}.

On $\widetilde{\mathrm{BZ}}_C$, the relation between the total projection and the rank-2 real vector bundle structure is not so obvious. Let us see how it unfolds for these complex eigenvectors. On $\widetilde{\mathrm{BZ}}_C$, the two eigenvalues correspond to the following block in the canonical form of $H(\mathbf k,\lambda)$:
\begin{equation}
    \begin{pmatrix}
       |\phi\rangle , &
        |\phi\rangle^\ast
    \end{pmatrix}
    \begin{pmatrix}
        \omega& 0\\
        0 & \omega^\ast
    \end{pmatrix}
    \begin{pmatrix}
         \langle\widetilde\phi|\\
        \langle\widetilde\phi|^\ast
    \end{pmatrix},
\end{equation}
where $|\phi\rangle=|\phi_+\rangle$ corresponds to the eigenvalue on the lower half complex plane and $|\phi\rangle^\ast=|\phi_-\rangle$ corresponds to the upper half complex plane partner. The two eigenvectors are biorthogonal: $\langle\widetilde\phi|\cdot|\phi \rangle^\ast=\langle\widetilde\phi|^\ast\cdot|\phi \rangle=0$ and $\langle\widetilde\phi|\cdot|\phi \rangle=\langle\widetilde\phi|^\ast\cdot|\phi \rangle^\ast=1$. Now the eigenvectors are complex. We decompose them into their real and imaginary parts: $|\phi \rangle=|\psi_R\rangle+i|\psi_I\rangle,\ \langle\widetilde\phi|=\langle\widetilde\psi_R|+i\langle\tilde\psi_I|$. There exist a series of relations between these vectors. First of all, the vectors $|\psi_R\rangle$ and $|\psi_I\rangle$ are linearly independent, and so are $\langle\widetilde\psi_R|$ and $\langle\widetilde\psi_I|$. Secondly, due to the biorthogonal relations, we can verify these real vectors are also biorthogonal:
$\langle\widetilde \psi_\alpha|\psi_\beta\rangle=\delta_{\alpha\beta}/2$. The sum projection for this pair of complex eigenvalues is
\begin{align}\label{eq:Psumrealim}
P_{\textrm{sum}}(\mathbf k,\lambda)=&|\phi(\mathbf k,\lambda)\rangle\langle\widetilde\phi(\mathbf k,\lambda)|+|\phi(\mathbf k,\lambda)\rangle^\ast\langle\widetilde\phi(\mathbf k,\lambda)|^\ast\nonumber\\
=&2\left[|\psi_R(\mathbf k,\lambda)\rangle\langle\widetilde\psi_R(\mathbf k,\lambda)|+|\psi_I(\mathbf k,\lambda)\rangle\right.\nonumber\\
&\left.\times\langle\widetilde\psi_I(\mathbf k,\lambda)|\right],\quad\textrm{for $(\mathbf k,\lambda)\in \widetilde{\mathrm{BZ}}_C$}.
\end{align}
This is exactly a projection operator into the real $2$-dimensional subspace of $\mathbb R^N$ spanned by $|\psi_R\rangle$ and $|\psi_I\rangle$. So in the situation of complex conjugate eigenvalues, the sum projection still defines a map from $\widetilde{\mathrm{BZ}}_C$ to  the space of two dimensional subspaces of $\mathbb R^N$:
\begin{equation}
    \widetilde{\mathrm{BZ}}_C\to \mathrm{Gr}_2(\mathbb R^N):\quad (\mathbf k,\lambda)\mapsto\textrm{img } P_{\textrm{sum}}(\mathbf k,\lambda).
\end{equation}
Again the above map is continuous on $ \widetilde{\mathrm{BZ}}_C$. Albeit the two eigenvectors are complex, their sum projection is projecting every real vector in $\mathbb R^N$ into the subspace spanned by the real and the imaginary parts of the eigenvector, i.e., $V_\lambda(\mathbf k)$. Notice that it does not matter which eigenvector is chosen to take the real and the imaginary parts, since the corresponding components of $|\phi\rangle$ and $|\phi\rangle^\ast$ span the same real $2$-dimensional subspace.  

Combining the two cases, the sum projection operator always gives a continuous map from non-degenerate regions of $\widetilde{\mathrm{BZ}}$ to the Grassmannian $\mathrm{Gr}_2(\mathbb R^N)$, leading to a continuous $V_\lambda(\mathbf k)$ for these regions. In the next step, we will show that the sum projection is also continuous on degenerate regions of $\widetilde{\mathrm{BZ}}$. Therefore the map to $\mathrm{Gr}_2(\mathbb R^N)$ (also $V_\lambda(\mathbf k)$) is continuous throughout the whole phase transition process. 

\emph{Continuity of $V_\lambda(\mathbf k)$ at EPs}: We first prove $(\mathbf k,\lambda)\mapsto P_{\textrm{sum}}(\mathbf k,\lambda)$ is also continuous over every degenerate point. Then using the continuity of the map $P_{\textrm{sum}}(\mathbf k,\lambda)\mapsto\textrm{img } P_{\textrm{sum}}(\mathbf k,\lambda)$ of constant-rank projection operators, we prove that the map from $\widetilde{\mathrm{BZ}}$ to $\mathrm{Gr}_2(\mathbb R^N)$ is continuous. This can be represented as the following composition map:
\begin{equation}
     \widetilde{\mathrm{BZ}}\to \mathrm{Gr}_2(\mathbb R^N):\quad (\mathbf k,\lambda) \mapsto P_{\textrm{sum}}\mapsto \textrm{img } P_{\textrm{sum}}.\label{eq_mgr}
\end{equation}

\begin{figure}
    \centering
    \includegraphics[width=0.8\linewidth]{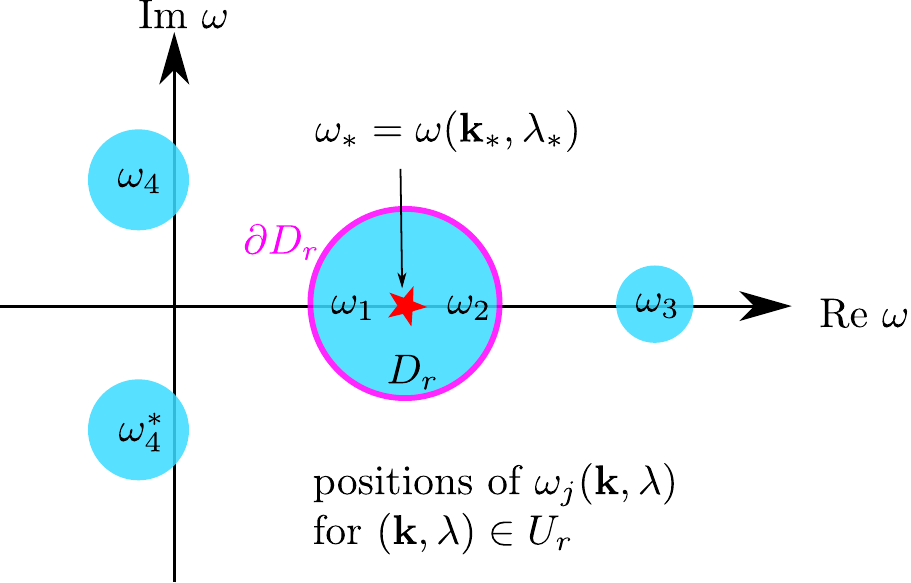}
    \caption{How to choose the enclosing disc $D$ on the complex plane of spectrum. The blue discs represent the approximate region of all eigenvalues $\omega_j$ for $\mathbf k,\lambda$ in a neighbourhood of the EP.}
    \label{fig_profill}
\end{figure}

At the degenerate point, it is not convenient to use the language of eigenvectors, as they become singular. However, the sum projection is still well defined, via the integral expression Eq.~\eqref{eq_pjo2}. We denote the degenerate point as $(\mathbf k_\ast,\lambda_\ast)$ and the corresponding eigenvalue at this point as $\omega_\ast\in \mathbb R$. Around $(\mathbf k_\ast,\lambda_\ast)$, we always denote the two eigenvalues as $\omega_1(\mathbf k,\lambda)$ and $\omega_2(\mathbf k,\lambda)$, regardless of whether they are real or a conjugate pair, or even equal. Since the eigenvalues are always  continuous in $H$ \cite{kato2013perturbation}, for any (open) disc region $D_r$ of radius $r$ centered at $\omega_\ast$, there is a neighbourhood $U_r$ of $(\mathbf k_\ast,\lambda_\ast)$ such that for all $(\mathbf k,\lambda)\in U_r$, $\omega_1(\mathbf k,\lambda),\omega_2(\mathbf k,\lambda)\in D_r$. As the two bands do not touch any other remote bands during the transition, we can let $U_r$ and $D_r$ small enough such that all other eigenvalues of $H(\mathbf k,\lambda)$ are outside $\overline D_r$ for all $(\mathbf k,\lambda)\in U_r$. 
With these definitions, we can safely evaluate $P_{\textrm{sum}}(\mathbf k,\lambda)$ by choosing $\mathcal D=D_r$:
\begin{equation}
   P_{\textrm{sum}}(\mathbf k,\lambda)=\frac{1}{2\pi i}\int_{\partial D_r}\left[z-H(\mathbf k,\lambda)\right]^{-1}dz,\,\forall (\mathbf k,\lambda)\in U_r.\label{eq_epttalP}
\end{equation}

We prove that Eq.~\eqref{eq_epttalP} is a continuous map at $(\mathbf k_\ast,\lambda_\ast)$. This step is similar to the proof of continuity for non-Hermitian spectral flattening (Ref.~\onlinecite{yang2023homotopy} Appendix.~G). Denote $A=z-H(\mathbf k_\ast,\lambda_\ast)$ and $B=z-H(\mathbf k_2,\lambda_2)$ with $(\mathbf k_2,\lambda_2)\in U_r$. By choosing $(\mathbf k_2,\lambda_2)$ close enough to $(\mathbf k_\ast,\lambda_\ast)$, we can let $\delta=||A-B||$ arbitrarily small. Then the difference between their reverse is
\begin{align}
    ||A^{-1}-B^{-1}||=&||B^{-1}(B-A)A^{-1}||\nonumber\\
    \le& ||B^{-1}||\cdot ||(B-A)|| \cdot||A^{-1}||\nonumber\\
    =&||B^{-1}||\cdot ||A^{-1}||\delta.\label{eq_rvd}
\end{align}
As $A$ invertible for $z\in\partial D_r$, $||A^{-1}||$ is a continuous bounded function in $z$. The value of $||B^{-1}||$ is estimated as below:
\begin{align}
    ||B^{-1}||=&||A^{-1}+B^{-1}(A-B)A^{-1}||\nonumber\\
\le&||A^{-1}||+||B^{-1}||\cdot||A^{-1}||\delta,\\
\rightarrow ||B^{-1}||\le&\frac{||A^{-1}||}{1-||A^{-1}||\delta}.\label{eq_rve}
\end{align}
Combining Eq.~\eqref{eq_rvd} and Eq.~\eqref{eq_rve}, we obtain the bounding equation for the difference between resolvent:
\begin{equation}
     ||A^{-1}-B^{-1}||\le \frac{||A^{-1}||^2\delta}{1-||A^{-1}||\delta}.
\end{equation}
The right-hand-side goes to zero as $\delta\to 0$. To estimate the integral Eq.~\eqref{eq_epttalP}, we notice that $||A^{-1}||/(1-||A^{-1}||\delta)$ is a continuous function on $\partial D_r$, as long as $\delta$ is chosen to be small enough. We may denote its supremum as $w$. Then the difference between the total projections is bounded by
\begin{equation}
    ||P_{\textrm{sum}}(\mathbf k_2,\lambda_2)-P_{\textrm{sum}}(\mathbf k_\ast,\lambda_\ast)||\le wr\delta,
\end{equation}
for all $(\mathbf k_2,\lambda_2)$ close enough to $(\mathbf k_\ast,\lambda_\ast)$. So this proves the continuity of the projection $P_{\textrm{sum}}(\mathbf k,\lambda)$ at the degenerate points of the two relevant bands. Together with what we have already shown, $P_{\textrm{sum}}(\mathbf k,\lambda)$ is continuous over the whole region $\widetilde{\mathrm{BZ}}=\mathrm{BZ}\times [0,1]$.

The last thing we need to verify in Eq.~\eqref{eq_mgr} is that $\textrm{img } P_{\textrm{sum}}$ is continuous in $P_{\textrm{sum}}$. 
It suffices to check the continuity at a neighbourhood for arbitrary projection operator $P$.
First, note that a continuous $P(\mathbf k,\lambda)$ has a fixed rank ($\textrm{rank }P=\textrm{tr }P$).
Consider the space $O^{(2)}$, the space of rank-$2$ projection operators in $\mathbb R^N$, and taking an arbitrary point $P\in O^{(2)}$.
We build a map from $U$, a neighbourhood of $P$ in $O^{(2)}$, to $\mathrm{St}_2(\mathbb R^N)$, the Stiefel manifold \cite{milnor1974characteristic}.  
We can pick two fixed vectors $v_1,v_2\in \mathbb R^N$ such that $Pv_1$ and $Pv_2$ are linearly independent. Then there is a neighbourhood $U$ of $P$ such that $P'v_1$ and $P'v_2$ are independent for all $P'\in U$. This defines a continuous map from $U$ to $\mathrm{St}_2(\mathbb R^N)$. Using the canonical map $\mathrm{St}_2(\mathbb R^N)\to \mathrm{Gr}_2(\mathbb R^N)$, the map $P_{\textrm{sum}}\mapsto\textrm{img } P_{\textrm{sum}}$ is continuous on $U$. 
This completes the proof of continuity of $\widetilde{\mathrm{BZ}}\to \mathrm{Gr}_2(\mathbb R^N)$.

Recall $\widetilde{\mathrm{BZ}}=\mathrm{BZ}\times[0,1]$.
At each $\lambda\in[0,1]$, we have a real vector bundle $F_{\mathbb R,\lambda}$ over $\mathrm{BZ}$ defind by $V_\lambda(\mathbf k)$ with Euler number denoted by $\chi_\lambda$.
The map $\widetilde{\mathrm{BZ}}\to \mathrm{Gr}_2(\mathbb R^N)$ is an homotopy between $\mathrm{BZ}\times\{0\}\to \mathrm{Gr}_2(\mathbb R^N)$ and $\mathrm{BZ}\times\{1\}\to \mathrm{Gr}_2(\mathbb R^N)$, which induces an isomorphism between two vector bundles at $\lambda=0$ and $\lambda=1$.
Therefore, we have $\chi(F_{\mathbb R,\lambda=0})=\chi(F_{\mathbb R,\lambda=1})$: the Euler number of the real vector bundle $F^{RI}$ spanned by $|\psi_R(\mathbf k,\lambda=1)\rangle,|\psi_I(\mathbf k,\lambda=1)\rangle$ is equal to that of the initial symmetry-preserving real band pair. 

\subsubsection{More details on realification}\label{sc_rlf}

% To prove $|\chi_{\lambda=1}|=|C_\pm|$, independent of the shape of the BZ, we employ the fact that the Chern class of a complex line bundle is equal to the Euler class of its underlying rank-2 real vector bundle. 
The eigenstate $|\phi(\vk)\rangle$ at $\lambda=1$ spans a complex line bundle $E_\mathbb C$. At each $\mathbf k$, this bundle gives an one dimensional complex vector space $\widetilde V_{\mathbb C^1}(\mathbf k)$. An one dimensional complex vector space can always be regarded as a two-dimensional real vector space $\widetilde V_{\mathbb R^2}(\mathbf k)$, called its realification \cite{bott1982differential}. This $\widetilde V_{\mathbb R^2}(\mathbf k)$ gives an underlying rank-two real vector bundle $E_{\mathbb R}$ corresponding to $E_\mathbb C$, for which we have $\chi(E_\mathbb R)=C(E_\mathbb C)=C_+$ \cite{milnor1974characteristic,bott1982differential}. 

There is a tiny difference in the realification $E_{\mathbb R}$ defined by mathematicians and the realification $F^{RI}$ that we used in the main text. Mathematicians realify bundles in the complex coefficients, while in the main text we realify bundles in the basis. We show that the two construction are isomorphic and do not cause any trouble in topology. So $E_\mathbb R$ and $F^{RI} $ share the Stiefel-Whitney and Euler classes. 
The isomorphism can be built as follows. 
For each $\mathbf k$, we pick $|\phi_+(\mathbf k)\rangle$ and expand it as $|\psi_R(\mathbf k)\rangle+i|\psi_I(\mathbf k)\rangle$. 
A complex vector $(a+ib)|\phi_+(\mathbf k)\rangle\in \widetilde V_{\mathbb C^1}(\mathbf k), a,b\in \mathbb R$ is first mapped to a vector $(a,b)\in \widetilde V_{\mathbb R^2}(\mathbf k)$ according to the definition of $E_\mathbb R$. Then we map this vector in $\widetilde V_{\mathbb R^2}$ to $a|\psi_R(\mathbf k)\rangle-b|\psi_I(\mathbf k)\rangle$ in span$\{|\psi_R(\mathbf k)\rangle,|\psi_I(\mathbf k)\rangle\}=V_1(\mathbf k)$. 
We can verify that this map is $\mathbb{R}$-linear, invertible,  and does not depend on the phase chosen for $|\phi_+(\mathbf k)\rangle$ (in fact, this map is simply $c\ket{\phi_+(\mathbf k)}\mapsto \text{Re } c\ket{\phi_+(\mathbf k)}$, where Re is globally well-defined).
So this is indeed a vector bundle isomorphism. 

Note that in this section, we do not need any topological property of the BZ torus apart from compactness. Therefore the Chern-Euler duality applies to Euler bands and Chern bands on any two-dimensional compact surfaces.

\subsection{Wilson loop results for the dual transition}\label{sc_wl}

We show how to use Wilson loop to find out the topological invariants and interpret the dual transition.

\begin{figure*}
    \centering
    \includegraphics[width=0.8\linewidth]{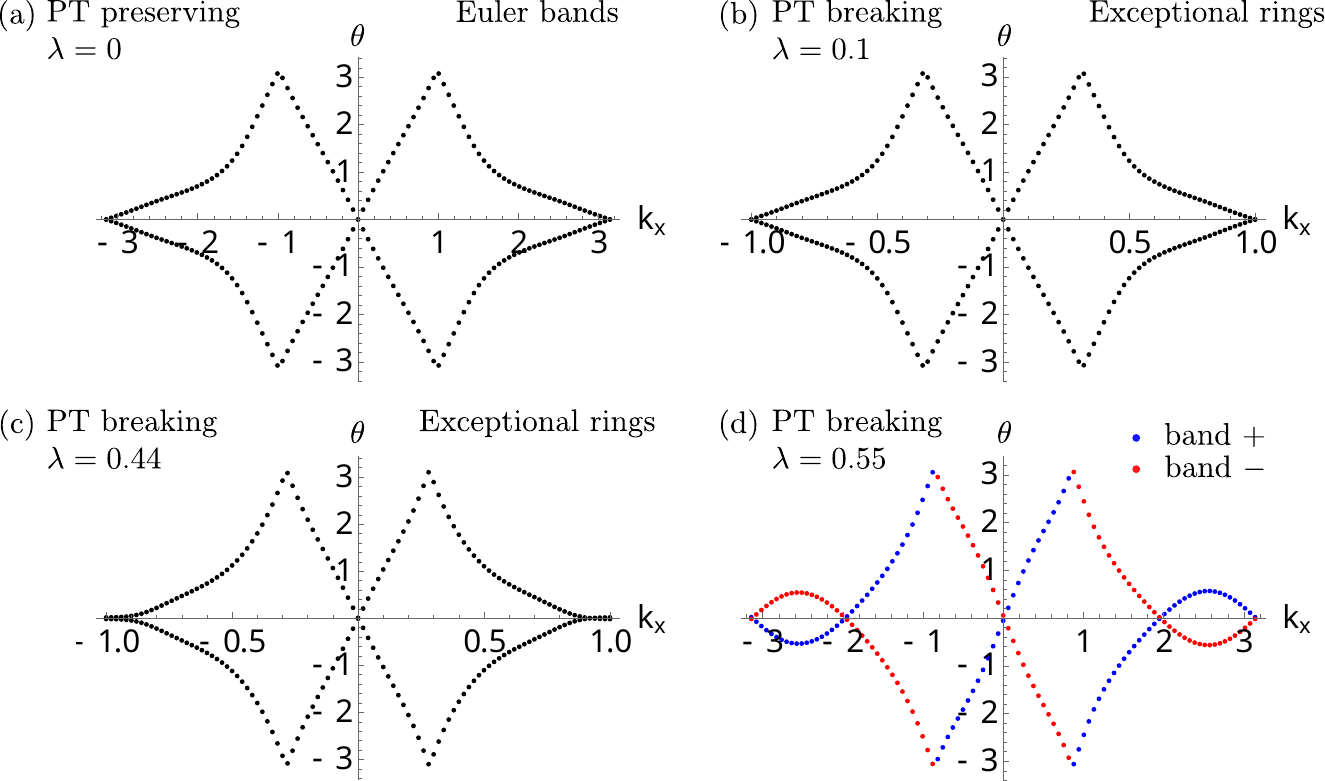}
    \caption{Wilson loop phases during the spontaneous Chern-Euler transition. (a) The Euler bands in the symmetry-preserving regime. The Wilson loop operator has two complex conjugate eigenvalues. Their phases are going through $\pm 2\chi\pi$ winding. (b)-(c) The Wilson loops correspond to the summation projection operator of the two bands going through the SSB. Even exceptional rings appear, the projection operator is continuous and deform adiabatically from the initial Euler bands. (d) The Chern bands in the SSB regime. Each Chern band possesses its own Wilson loop operator, whose action is a complex number inside the band. The phase of this complex number does a $2C\pi$ rotation.}
    \label{fig_wsEL}
\end{figure*}

First, we construct the Wilson loop operator from successive action of projection operators. We take the loop in the direction of $k_y$ along fixed $k_x$. The start and end point is chosen to be $(k_x,-\pi)=(k_x,\pi)$. The operator is given by
\begin{align}
O_w(k_x)=\lim_{\delta\to 0}\prod_{n}&P(k_x,\pi)P(k_x,\pi-\delta)\dots \nonumber\\
&P(k_x,\pi-n\delta)\dots P(k_x,-\pi).\label{eq_wlp}
\end{align}
The projection operator can be obtained from the $H$ directly. As we have seen in Eq.~\eqref{eq_pjo}, the projection operator in the non-Hermitian regime still only relies on the Hamiltonian matrix and is $\mbox{GL}_n(\mathbb C)$-gauge-independent or $\mbox{GL}_n(\mathbb R)$-gauge-independent. To obtain it, we do a Jordan decomposition $H=V^{-1}D_JV$. The matrix $D_J$ is a block-diagonal matrix with Jordan blocks corresponding to eigenvalues of $H$. To find out the projection operator $P_j$ for the $j$-th eigenvalue $\omega_j$ (it may be $n$-fold degenerate), we replace the $n\times n$ (Jordan) block corresponding to $\omega_j$ in $D_J$ by an $n$-dimensional identity matrix, and set all other elements of $D_J$ to zero. Via this, we obtain an $N\times N$ matrix $e_j$. Then $P_j$ is given by $P_j=V^{-1}e_jV$. This method has the merit of gauge independence and one can verify this is indeed a well-defined projection $P^2_j=P_j$. 

For the Hermitian Euler bands, we take the projection operator to be the total projection on the Euler bands $P(\mathbf k)=P_{\textrm{sum}}(\mathbf k)=|\psi_1(\mathbf k)\rangle\langle \psi_1(\mathbf k)|+|\psi_2(\mathbf k)\rangle\langle \psi_2(\mathbf k)|$, numerically obtained via the methods in the last paragraph. The loop operator sandwiched between the eigenstates defines a linear transformation matrix:
\begin{equation}
    W_{ij}(k_x)=\langle \psi_i(k_x,-\pi)|O_w(k_x)|\psi_j(k_x,-\pi)\rangle, i,j\in \{1,2\}.\label{eq_wleu}
\end{equation}
This matrix can be uniquely polar decomposed into the product of an orthogonal matrix $U$ and a positive definite symmetric matrix $S$ (in practice, we can do singular value decomposition $W=V_1\Sigma V_2^{-1}$ and set the diagonal matrix to identity to obtain $U=V_1V_2^{-1}$):
\begin{equation}
     W_{ij}(k_x)=\sum_l U_{il}(k_x)S_{lj}(k_x).\label{eq_pdcom}
\end{equation}
The eigenvalues of the $2\times 2$ orthogonal matrix $U_{ij}(k_x)$ are a pair of complex conjugate numbers $\{\exp[i\theta(k_x)],\exp[-i\theta(k_x)]\}$. As a periodic function of $k_x$, the phase $\theta(k_x)$ increases by an integer multiple of $2\pi$ when $k_x$ goes from $-\pi$ to $\pi$. This integer is the Euler number $\chi$ \cite{PhysRevB.102.115135}. The result for our 3-band model $H_{\mathrm{Eu}}(\mathbf k)$ is shown in Fig.~\ref{fig_wsEL}a.

In Sec.~\ref{sc_dpf2}, we showed that the oblique projection $P_{\textrm{sum}}$ is continuous even in the regime with exceptional rings. With this property, it is possible to study the corresponding Wilson loop  of $P_{\textrm{sum}}$ for $H_{0<\lambda<1}$. The operator $P_{\textrm{sum}}$ are constructed by Jordan-decomposing $H_\lambda$, replacing the block related to the pairs of bands by identity matrix and all other elements by 0. The operator $O_w$ is constructed as Eq.~\eqref{eq_wlp}. The only difference from Hermitian regime is that the matrix element \eqref{eq_wleu} is slightly harder, since we may touch exceptional rings. The strategy is to first find out the two eigenvectors of $P_{\textrm{sum}}(k_x,-\pi)$ corresponding to the unit eigenvalue. Then we orthogonormalise these two eigenvectors into $v_1(k_x)$ and $v_2(k_x)$. The $2\times 2$ matrix $W_{ij}(k_x)$ is now obtained by replacing $|\psi_{1,2}(k_x,-\pi)\rangle$ by $v_{1,2}(k_x)$ in Eq.~\ref{eq_wleu}:
\begin{equation}
    W_{ij}(k_x)=\langle v_i(k_x)|O_w(k_x)|v_j(k_x)\rangle, i,j\in \{1,2\}.
\end{equation}
Doing a similar polar decomposition as Eq.~\ref{eq_pdcom}, we extract the eigenphases of the orthogonal component.
Their phase winding are plotted in Fig.~\ref{fig_wsEL}b-c. We can see the topology is preserved through the spontaneous symmetry breaking process, in spite of the presence of exceptional rings.

The Wilson loop approach is also applicable for Chern bands. For a Chern band $|\psi_j(\mathbf k)\rangle$, we take the projection operator in Eq.~\eqref{eq_wlp} to be $P(\mathbf k)=P_j(\mathbf k)=|\widetilde\psi_j(\mathbf k)\rangle\langle\psi_j(\mathbf k)|$. This matrix is numerically obtained by Jordan decompositon $H=V^{-1}D_JV$ and replacing the Jordan block in $D_J$ corresponding to the band $j$ by $1$ and all other elements by $0$. The transformation matrix $W$ is now a complex number $c\ne 0$. The polar decomposition simply gives the phase of $c$, whose winding when $k_x$ is swept from $-\pi$ to $\pi$ gives the Chern number. We give this results for $H_{\lambda=0.55}$ in Fig.~\ref{fig_wsEL}d. Since now the two bands $+,-$ are separated by a dissipation gap, each of their projections $P_\pm$ gives a Wilson loop phase winding $\pm 4\pi$. Putting them together, we obtain a pattern joint pattern which shares the same topology as the original Euler phase but with geometric distinctions.

By comparing results in Fig.~\ref{fig_wsEL}, we see when the symmetry is completely broken and an imaginary eigenvalue gap is opened, the original Euler topology is inherited by each individual Chern bands. There are some geometric differences between the Wilson loops corresponding to $P_{\textrm{sum}}$ and $P_+,P_-$, due to basis choices. Nevertheless, they do not alter the topological information.

\subsection{Covariant connection VS Hermitian ``connection''}\label{sc_cpcn}

We clarify the issues with Berry connection and curvature in this section. We start from basic introduction to vector bundles and the connection on it. In this section, we use the term ``gauge transformation'' to refer to basis transformation inside a vector bundle, as in many physical literature.

We have a wave function $|\psi_j(\mathbf k)\rangle$ depending on parameters $\mathbf k=(k_1,k_2,\dots)$. The wave functions different by a scalar factor represent the same physical state. In this sense, each physical state corresponds to an one-dimensional vector space and several distinct states $|\psi_1(\mathbf k)\rangle,|\psi_2(\mathbf k)\rangle,\dots$ can form a higher dimensional vector space at $\mathbf k$. The parameters $\mathbf k$ define a base space where the vector spaces grow upon. A natural language to describe such structure is vector bundle, if the vector spaces are continuous with respect to $\mathbf k$. The dimension of the vector space is called the rank of the vector bundle. A common physical situation is the wave functions coming out of the eigenstates of a Hamiltonian matrix $H(\mathbf k)$. The vector space spanned by each eigenstate is usually continuous in $\mathbf k$ when there is no level crossings, or nontrivial loops in the base space. In this situation, each band corresponds to a rank-1 vector bundle, or a line bundle. When a singularity or discontinuity of a single-level wave function takes place, we may look at the combined vector space including all levels associated with the singularity. Usually the combined vector space is continuous and permits a vector bundle description. This has also been seen in our proof of the Chern-Euler duality. 

The singularities of level crossings necessitate gauge transformations. For example, in 2D systems, the eigenstate wave functions are not continuous at the Dirac node between two bands, so is the individual vector space spanned by each eigenstate. However, we can always perform a $\mbox{SO}(2)$ gauge transformation on the two eigenstates to explicitly see the sum vector space spanned by them is continuous at the Dirac point \cite{bouhon2020non}. Due to this continuity, it is possible to describe the two eigenstates near the Dirac node using the language of rank-2 vector bundles .

Once we have a well-defined vector bundle, geometric quantities naturally appear. A geometric connection tells us how to take directional derivatives for vector fields on a vector bundle, in other words, how to compare the vector spaces at different points of the base space. As an intrinsic property of the vector bundle, the connection should not rely on which basis is chosen. Therefore the geometric connection is covariant under a gauge transformation on the eigenstates,.

In Hermitian systems, we usually only need to perform orthogonal or unitary gauge transformations, as the example around Dirac points. In non-Hermitian systems, this may not be enough. An important example is the EP, where the eigenstates overlap and a unitary gauge transformation cannot restore the correct smooth basis. Another example is the braiding of eigenvalues. There the eigenstates swap. Since the eigenstates are not orthogonal to each other, there may be no guarantee that the swap  matrix of the eigenstates is unitary or orthogonal. Instead, we have to replace the usual orthogonal/unitary transformations by general linear transformations $\mbox{GL}_N(\mathbb R),\mbox{GL}_N(\mathbb C)$.

Motivated by the above requirements, a \emph{covariant connection} is defined by \cite{yang2023homotopy}:
\begin{equation}
   \nabla_a|\psi_j(\mathbf k)\rangle\equiv
   \partial_a|\psi_j(\mathbf k)\rangle=-i\sum_{j'} \mathcal A^{j'j}_{a}(\mathbf k)|\psi_{j'}(\mathbf k)\rangle,\label{eq_dfcc}
\end{equation}
where $|\psi_j\rangle$ is the right eigenvector of the $j$-th band. We denote the corresponding biorthogonal left eigenvector as $\langle\widetilde\psi_j|$. The first equation is the definition of the geometric connection $\nabla$, while the second equation is to expand $ \partial_a|\psi_j(\mathbf k)\rangle$ in terms of the eigenbasis. The coefficients form the connection matrix $\mathcal A^{j'j}_{a}$. For diagonalizable matrices,  we can use the biorthogonal relation $\langle \widetilde\psi_i(\mathbf k)|\psi_j(\mathbf k)\rangle=\delta_{ij}$ [for non-diagonalizable matrices, we need to include generalised eigenvectors. A better way is to use the basis-independent expression Eq.~\eqref{eq_bccdf}]. The connection matrix is
\begin{equation}
    \mathcal A^{ij}_{a}(\mathbf k)= i\langle \widetilde\psi_i(\mathbf k)|\partial_a\psi_j(\mathbf k)\rangle.\label{eq_bcm}
\end{equation}
The covariance under gauge transformation is computed directly from Eq.~\eqref{eq_dfcc}. If we choose another basis $|\psi'_j(\mathbf k)\rangle=\sum_i |\psi_i(\mathbf k)\rangle V_{ij}(\mathbf k)$ on $F$ [the biorthongoal left eigenvectors transform as $\langle\widetilde\psi_j(\mathbf k)|\to \sum_i\langle\widetilde\psi_i(\mathbf k)|V^{-1}_{ji}(\mathbf k)$], the new connection matrix is related to the Berry connection matrix through the Leibniz rule:
\begin{equation}
    \mathcal A'_{a}(\mathbf k)=V^{-1}(\mathbf k)\mathcal A_{a}(\mathbf k)V(\mathbf k)+iV^{-1}(\mathbf k)\partial_a V(\mathbf k).\label{eq_bctr}
\end{equation}
This is indeed the rule required to be a geometric connection \cite{milnor1974characteristic}. This transformation will guarantee the invariance of intrinsic topological and geometrical quantities for generic multi-band situations, as we will see in Sec.\ref{sc_nabc}. Most importantly, the covariance enables a sum vector bundle description at EPs, which will be explained in the Sec.~\ref{sc_sqm}.

The covariant connection matrix satisfies well-defined geometric properties. However sometimes it requires more work to compute, since the left eigenstates are the inverse of the right eigenstates (all right eigenvectors form an invertible matrix and the biorthogonal left eigenvectors is the inverse of this matrix). One may ask if there are situations where we only need to find out right eigenstates. Simplification happens when all gauge transformations can be reduced from general linear groups to unitary $\mbox{U}(N)$ or orthogonal $\mbox{O}(N)$ groups. A particular example is that the bands we study are spectrally isolated from each other \cite{PhysRevLett.120.146402,yang2023homotopy}. There the gauge transformation is reduced to a diagonal unitary or orthogonal matrix. We can use the following \emph{Hermitian ``connection'' matrix} to capture topological properties under this condition:
\begin{equation}
     A^{ij}_{a}(\mathbf k)= i\langle \psi_i(\mathbf k)|\partial_a\psi_j(\mathbf k)\rangle,
\end{equation}
where now $\langle \psi_j(\mathbf k)|$ is just the complex conjugate of $|\psi_j(\mathbf k)\rangle$ and the states are normalized according to $\langle \psi_j(\mathbf k)|\psi_j(\mathbf k)\rangle=1$. Since the operator $i\partial_a$ is Hermitian, the quantity $A^{ij}_{a}$ is also an Hermitian matrix.

Under a change of basis $|\psi'_j(\mathbf k)\rangle=\sum_i |\psi_i(\mathbf k)\rangle V_{ij}(\mathbf k)$, the matrix $A^{ij}_{a}$ transforms as 
\begin{equation}
    A'_{a}(\mathbf k)=V^{\dagger}(\mathbf k)A_{a}(\mathbf k)V(\mathbf k)+iV^{\dagger}(\mathbf k)\partial_a V(\mathbf k).
\end{equation}
The above transformation is not covariant for generic linear transformations, but is so for $V\in \mbox{U}(N)$ or $V\in \mbox{O}(N)$ since for these matrices $V^\dagger=V^{-1}$. Hence if we study systems where only unitary or orthogoanl gauge transformations on the eigenstates are involved, the Hermitian ``connection'' matrix is also a well-defined geometric quantity. This is the situation when the system itself is Hermitian, or each band studied does not braid with or cross other bands via EPs.

To conclude, the Hermitian ``connection'' is directly applicable when studying the topology of bands without eigenvalue braidings or EPs, which is the situation of the 3-band models in the main text. However, for generic multi-band systems, albeit being easier to compute, the Hermitian ``connection'' matrix does not have an elegant geometric meaning as in Eq.~\eqref{eq_dfcc}. Moreover, if we want to study the total topology of several bands connected by EPs or bands braiding together, we must perform gauge transformations beyond the orthogonal and the unitary groups. The Hermitian ``connection'' matrix does not correspond to a geometric connection in these cases and an additional orthonormalization procedure on the eigenstates such as the one in the Appendix is required.  This is why we use quotation marks ``connection''. Nevertheless, since $A^{ij}_{a}(\mathbf k)$ has the merit of being Hermitian, it may have more physical relevance when discussing measurable quantities. We leave these discussions to future.

\subsection{Computations of the topological invariants}\label{sc_nabc}

Euler number and Chern number are the most interesting topological invariants in geometry as they can be given by integral of geometric quantities, i.e., through de Rham cohomology. The key ingredients are the invariant polynomials of curvature matrices.

With a given connection matrix, the curvature matrix is given by:
\begin{equation}
     \mathcal B^{ij}=\partial_x\mathcal A^{ij}_{y}-\partial_y\mathcal A^{ij}_{x}-i[\mathcal A_x,\mathcal A_y]^{ij},\label{eq_bc}
\end{equation}
where the last term is the commutator of the connection matrix. If we are choosing the covaviant Berry connection in this formula, the corresponding Berry curvature matrix is  also covariant under a change of basis:
\begin{equation}
    \mathcal B(\mathbf k)\to V^{-1}(\mathbf k)\mathcal B(\mathbf k) V(\mathbf k) \label{eq_trfcur}
\end{equation}

The Euler number and Chern number are computed according to the scheme in Refs.~\onlinecite{bouhon2020non} and \onlinecite{yang2023homotopy}. Chern classes correspond to $\mbox{GL}_N(\mathbb C)$-invariant polynomials of the curvature matrix \cite{milnor1974characteristic}. So the general linear gauge transformations in non-Hermitian systems are automatically included in computation of Chern number. Using the covariant Berry curvature matrix, the Chern number is given by \cite{yang2023homotopy}:
\begin{equation}
    C=\int\frac{1}{2\pi} \textrm{tr } \mathcal B (\mathbf k)d^2 k,\label{eq_cn}
\end{equation}
where the trace is taken with respect to the bands studied. Notice that due to the covariance Eq.~\eqref{eq_trfcur}, the above equation is independent of gauge choice, as required of being a topological number.

As discussed in Sec.~\eqref{sc_cpcn}, if the gauge transformation of the non-Hermitian system can be reduced to unitary groups, the Hermitian connection matrix is also a well defined quantity. We may similarly define an Hermitian Berry curvature and its transformation rule is given by 
\begin{align}
    &B^{ij}=\partial_x A^{ij}_{y}-\partial_y A^{ij}_{x}-i[A_x, A_y]^{ij},\\
    &B(\mathbf k)\to V^{\dagger}(\mathbf k)B(\mathbf k) V(\mathbf k)    
\end{align}
The above equation is only covariant for $V(\mathbf k)\in \mbox{U}(N)$. Specific to our 3-band models, since each of the Chern band is spectrally separated from other bands, the gauge transformations can be reduced to $\mbox{U}(1)$ transformations. So their Chern numbers are given by the integral of the Hermitian Berry curvature as well \cite{PhysRevLett.120.146402,yang2023homotopy}. This greatly simplifies the calculations and the measurements.

The Euler class is a $\mbox{SO}(N)$-invariant polynomial [not $\mbox{GL}_N(\mathbb R)$-invariant] in the curvature matrix. In the non-Hermitian settings, it will require more orthonormalization procedures to be obtained from integrals of curvature. Readers with more interests may check Ref.~\onlinecite{yang2023homotopy}. The Euler number exists when the vector bundle is real and orientable. In this situation, the curvature matrix of a metric-compatible connection under an orthonormal basis is skew-symmetric. The Euler form is the Pfaffian of this matrix. Specific to two Euler bands in 2D systems, this is the off-diagonal component of the Berry curvature
\begin{equation}
    \chi=\int \frac{1}{2\pi i}B^{12}(\mathbf k)d^2k.\label{eq_euf}
\end{equation}
One may verify the above expression is indeed invariant under a $\mbox{SO}(2)$ gauge transformation on bands 1 and 2. As we are studying Hermitian Euler bands, we simply use the Hermitian Berry curvature here.

There are some cautions in choosing the wave function gauge for the Euler form. The Euler form is the off-diagonal elements of the non-Abelian Berry curvature. Under a gauge transformation $|\psi_1\rangle\to |\psi_1\rangle,|\psi_2\rangle\to -|\psi_2\rangle$, the off-diagonal Berry curvature changes its sign. 
This simply reflects that the Euler form is not invariant under $\mbox{O}(2)$ transformations that include a mirror reflection. Although the Euler number is topologically quantized, the integral Eq.~\eqref{eq_euf} is equal to it only if we choose a consistent $\mbox{SO}(2)$ gauge over the BZ. The gauge is chosen following Ref.~\onlinecite{bouhon2020non} . There are even number of Dirac nodes inside the BZ (double the Euler number), we pairwise connect them via Dirac strings (this is to say we draw lines to pair them in the BZ). Outside the Dirac strings, the wave functions $|\psi_1(\mathbf k)\rangle$ and $|\psi_2(\mathbf k)\rangle$ can be written down continuously. And when we cross a Dirac string, we let the wave function different by a gauge 
transformation $|\psi_1\rangle\to -|\psi_1\rangle,|\psi_2\rangle\to -|\psi_2\rangle$. This will respect the topology of the Dirac nodes, going around which the wave function flips its sign. Under this gauge choice, although the wave function is not continuous throughout the BZ, the Euler form is. Then the integral of the Euler form gives the correct Euler number. 

The discrete numerical evaluation of the Euler number, together with the Chern number in the main text are done with the methods from Ref.~\onlinecite{DiscreBC}.

\subsection{Singularities of quantum geometry near EPs}\label{sc_sqm}

We are going to discuss the covariant (non-Abelian) curvature near EPs. We focus on the EPs where two bands cross.  Although the vector bundles spanned by the two eigenvectors are smooth, the covariant Berry curvature matrix under the eigenbasis can often be singular. The reason is that the transformation from a smooth basis of the vector bundles to the eigenbasis is always singular near EPs, because of eigenvector coalescence.

Let us understand this singularity. We go back to the definition of the Berry connection. We have two bands crossing each other through an EP. Without loss of generality, we denote them as band 1 and band 2. For our question, these two relevant band eigenvectors form a bundle $F$, which is a subbundle of the trivial vector bundle $\textrm{BZ}\times \mathbb R^N$. In order to look at the geometric connection $\nabla^F$ on $F$, we need to project the summation of $|\psi_{j'}\rangle$ in Eq.~\eqref{eq_dfcc} to $\textrm{span}\{|\psi_1\rangle,\psi_2\rangle\}$, i.e., restricting the summation to $j'=1,2$ \cite{yang2023homotopy}. The geometric meaning of Eq.~\eqref{eq_dfcc} is very simple, it tells us that the connection is defined to take the directional derivative by comparing the vectors in the trivial bundle $\textrm{BZ}\times \mathbb R^N$. In the trivial bundle, the vector space does not rotate or stretch with respect to the base manifold and thus comparison of vectors is unambiguous. When restricted to the subbundle $F$,  the connection $\nabla^F$ is the projection of the directional derivative in $\textrm{BZ}\times \mathbb R^N$ to $F$. We can thus rewrite the projected connection $\nabla^F$ in a more basis-independent form:
\begin{equation}
    \Gamma(T\textrm{BZ})\times \Gamma(F)\to \Gamma(F):\ \nabla^F_{\mathbf k}v(\mathbf k)=P_{\textrm{sum}}(\mathbf k)\partial_\mathbf k \iota[v (\mathbf k)],\label{eq_bccdf}
\end{equation}
where $\iota$ is induced by the inclusion map $F\to \textrm{BZ}\times \mathbb R^N$ and $\iota(v)$ therefore can be simply written as $v(\mathbf k)\in\mathbb R^N$ due to the trivial bundle structure. The symbol $\Gamma(\cdot)$ means sections of a vector bundle (they are vector fields along the BZ) and $T\textrm{BZ}$ represents the tangent bundle of the BZ. The inclusion map and the projection satisfy $P_{\textrm{sum}}\circ \iota=id_F $, where $id_F$ is the identity map on $F$.

From this definition, the geometric connection (not the connection matrix) does not depend on which basis we choose for $F$. The section $v$ in Eq.~\eqref{eq_bccdf} can be chosen to be any vector field on $F$ besides the eigenvectors. This independence is also reflected by directly calculating the connection matrix under another basis. 
Using a linear transformation on $|\psi_1\rangle,|\psi_2\rangle$, we can verify that Eq.~\eqref{eq_bctr} indeed agrees with Eq.~\eqref{eq_bccdf}.

From what we showed in Sec.~\ref{sc_dpf2}, the operator $P_{\mathrm{sum}}$ is continuous at EPs while $\iota$ is always continuous. So all operations in Eq.~\eqref{eq_bccdf} are continuous at  EPs. The geometric connection for the vector bundle spaned by $|\psi_1\rangle,\psi_2\rangle$ is therefore continuous. From the discussion in the last paragraph, there exists a smooth basis $v_1(\mathbf k),v_2(\mathbf k)$ spanning the fibers (vector spaces) of $F$ in the neighbourhood of the EP. The $2\times 2$ Berry connection matrix $\mathcal B_v(\mathbf k)$ under this basis is continuous, As we have shown in Sec.~\ref{sc_nabc}, the covariant  curvature matrix transforms as $V^{-1}\mathcal B_vV$ under a change of basis. Because the eigenvectors $|\psi_1\rangle,|\psi_2\rangle$ coalesce near the EP, their transformation $V(\mathbf k)$ from the smooth basis $v_1,v_2$ is not invertible at the EP. As a consequence, the matrix $V^{-1}(\mathbf k)$ diverges. This leads to a singularity of the curvature matrix under eigenbasis. A more straightforward understanding is that the biorthongal left eigenvectors must diverge near the EP if the right eigenvectors are chosen to be continuous. However, the true underlying geometric connection of the sum bundle $F$ is still continuous. The singularity only reflects the fact that the eigenvectors are not a good basis for the vector space they span when approaching EPs. 

We see that the singularity of curvature matrix is only an artifact from the divergence of linear transformations to eigenvectors near EPs. The geometric connection and curvature themselves are continuous. This discussion also justifies why we need to employ the covariant curvature matrix when studying the overall topological structure of multiple bands meeting at an EP, as a divergent $\mbox{GL}(\mathbb C)$ or $\mbox{GL}(\mathbb R)$ transformation will restore the continuity. In the end of this section, we remark that only the sum bundle $F$ is continuous at the EP, while the individual line bundle of each band is intrinsically singular at the EP. This is also seen as we cannot define the individual projection $P_1$ or $P_2$ via Eq.~\eqref{eq_pjo} when $\omega_1=\omega_2$. Only their sum $P_{\textrm{sum}}$ is a meaningful quantity.

\subsection{More examples for the Chern-Euler duality principle}

\begin{figure*}
    \centering
    \includegraphics[width=0.7\linewidth]{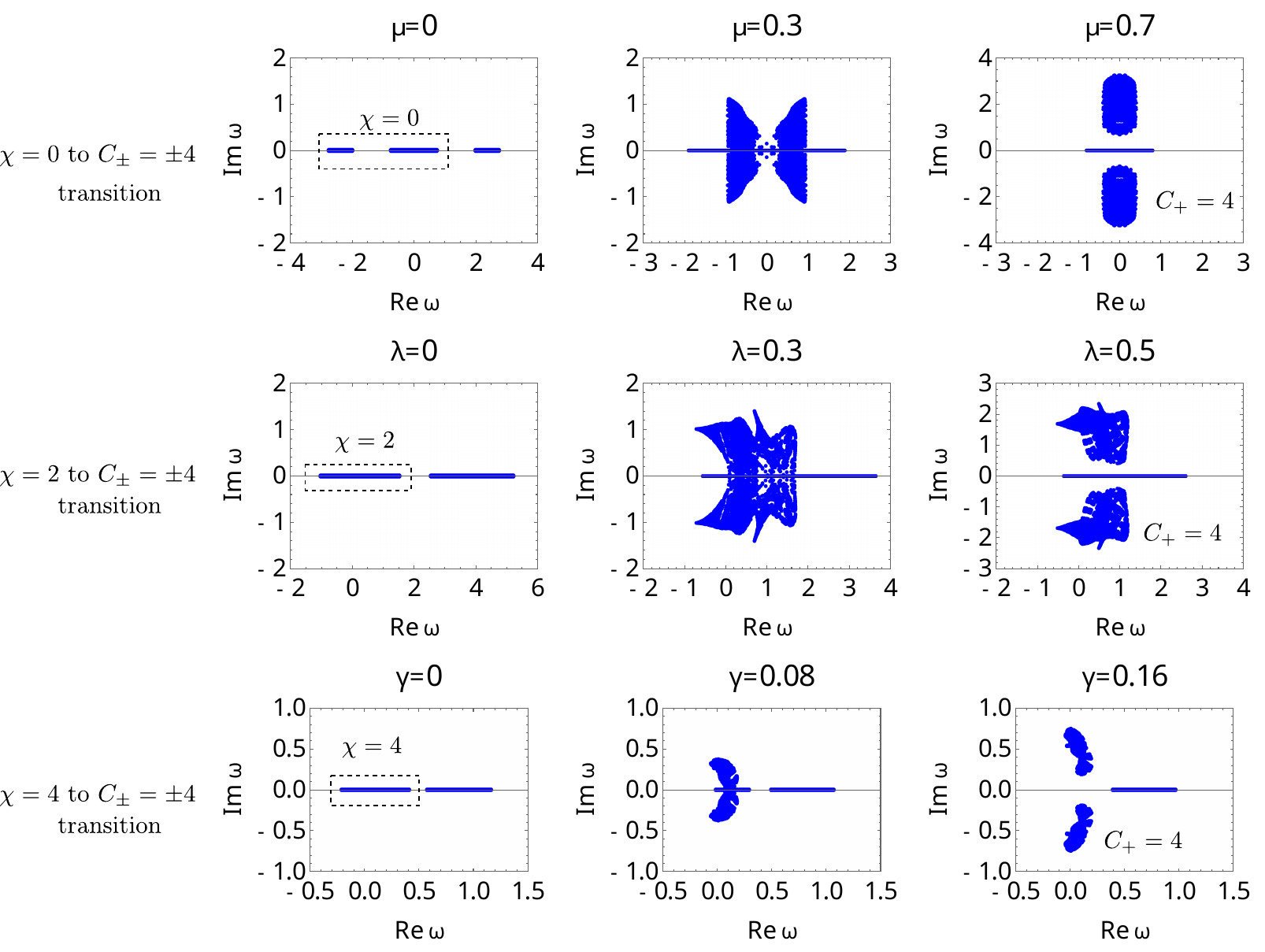}
    \caption{The spectra for transitions to complex bands of Chern numbers $C_\pm=\pm4$. The three rows correspond to spontaneous PT-symmetry breaking transitions to $C_\pm=\pm 4$ Chern bands from different initial real bands. During non-dual spontaneous PT-symmetry breaking, triple-band touchings must take place  when $|C_\pm|\ne|\chi|$, as reflected by the first two rows. A direct transition to the $C_\pm=\pm4$ bands from two real bands can take place when the initial model carry $\chi=4$, as in the last row. } 
    \label{fig_c4cases}
\end{figure*}

We give more examples for illustration of the general Chern-Euler duality principle in spontaneous PT-symmetry breaking transitions. 

For three-band PT-symmetric models, the next minimal examples are the transitions to a pair of $C_\pm=\pm 4$ complex bands. To realize such spontaneous PT-symmetry breaking, we employ an anti-Hermitian Hamiltonian $H_{|C|=4}$ whose explicit form is given by:
\begin{equation}
    H_{|C|=4}(\mathbf k) =\,
  \begin{pmatrix}
    0 & h_{12}(\mathbf k) & h_{13}(\mathbf k) \\
    -h_{12}(\mathbf k) & 0 & h_{23}(\mathbf k) \\
    - h_{13}(\mathbf k) & -h_{23}(\mathbf k) & 0
  \end{pmatrix}, 
\end{equation}
where the three functions are given by $h_{12}(\mathbf k)=0.4-0.5(\cos k_x+\cos k_y)-0.28[\cos(k_x+k_y)+\cos(k_x-ky)]$, $h_{13}(\mathbf k)=0.52[\cos(k_x-k_y)-\cos (k_x+k_y)]$, and $h_{23}(\mathbf k)=0.34(\cos k_x-\cos k_y)$. The complex band of this model on the lower half complex plane has Chern number $C_+=4$. The one on the upper half-complex plane carries $C_-=-4$ and the one on the real axis is a trivial band.

In order to provide a full illustration of the Chern-Euler duality principle, we also give an Hermitian PT-symmetric model $H_{\chi=4}$ with $\chi=4$ bands:
\begin{equation}
    H_{\chi=4}(\mathbf k) =\,
  \begin{pmatrix}
    f_{11}(\mathbf k) & f_{12}(\mathbf k) & f_{13}(\mathbf k) \\
    f_{12}(\mathbf k) & f_{22}(\mathbf k) & f_{23}(\mathbf k) \\
    f_{13}(\mathbf k) & f_{23}(\mathbf k) & f_{33}(\mathbf k)
  \end{pmatrix}. 
\end{equation}
The higher-Euler-number model requires a longer hopping range compared to the models in the main text. The functions in the Hamiltonian matrix are slightly more complicated. The more complicated band structure can be intuitively understood as more nodal points (number bound by the Euler number) have to appear between the two lower-energy bands. The diagonal components of $H_{\chi=4}$ are given by
\begin{align}
    f_{11}(\mathbf k)=&0.21+0.14(\cos k_x+\cos k_y)+0.08(\cos 3k_x+\nonumber\\&\cos 3k_y)-\sum_{\pm}[0.12\cos(2k_x\pm2k_y)+0.1\cos(k_x\nonumber\\
    &\pm 2k_y)+0.1\cos(2k_x\pm k_y)],\nonumber\\
    f_{22}(\mathbf k)=&0.29-0.26(\cos 2k_x+\cos 2k_y)+0.12\sum_\pm\cos(2k_x\nonumber\\
    & \pm 2k_y),\nonumber\\
    f_{22}(\mathbf k)=&0.5-0.18(\cos k_x+\cos k_y)+0.24(\cos 2k_x+\cos\nonumber\\
    & 2k_y)+0.1\sum_{\pm}[\cos(k_x\pm2k_y)+\cos(2k_x\pm k_y)].
\end{align}
The summation $\Sigma_\pm$ means that we need to sum all the terms corresponding to $+$ and $-$ signs. The off-diagonal components of $H_{\chi=4}$ similarly require longer-range hoppings and are given by the expressions:
\begin{align}
     f_{12}(\mathbf k)=&0.12[\cos(k_x-2k_y)+\cos(2k_x+k_y)]-0.12\nonumber\\&[\cos(2k_x-k_y)+\cos(k_x+2k_y)],\nonumber\\
    f_{13}(\mathbf k)=&0.16(\cos k_x-\cos k_y)+0.1(\cos 3k_y-\cos 3k_x),\nonumber\\
    f_{23}(\mathbf k)=&0.08[\cos(2k_x-2k_y)-\cos(2k_x+2k_y)]+0.14\nonumber\\&[\cos(k_x-2k_y)+\cos(2k_x-k_y)-\cos(2k_x+k_y)\nonumber\\
    &-\cos(k_x+2k_y)]+0.12[\cos(k_x+k_y)-\cos(k_x\nonumber\\
    &-k_y)].
\end{align}

We now use anti-Hermitian $H_{|C|=4}$ to spontaneously break the PT symmetry in the models $H_{\mathrm{trivial}},H_{\mathrm{Euler}}$ of the main text, as well as $H_{\chi=4}(\mathbf k)$ constructed above. This is achieved via the following interpolations:
\begin{align}
    H_\mu&=(1-\mu)H_{\mathrm{trivial}}+\mu H_{|C|=4},\nonumber\\
    H_\lambda&=(1-\lambda)H_{\mathrm{Euler}}+\lambda H_{|C|=4},\nonumber\\
    H_\gamma&=(1-\gamma)H_{\chi=4}+\gamma H_{|C|=4}.
\end{align}
According to the Chern-Euler duality principle, only the last spontaneous PT-symmetry breaking transition can take place with two bands, while the other two transitions involve triple-band touchings. This is indeed the case, as seen in Fig.~\ref{fig_c4cases}. The two non-dual transitions $|\chi|=0\to|C_\pm|=4,|\chi|=2\to|C_\pm|=4$ exhibit enforced triple-band touchings. Only the transition $|\chi|=4\to|C_\pm|=4$ takes place within two bands.

%TC:endignore

\end{document}